\documentclass[a4paper,11pt]{article}
\usepackage{jcappub} 

\usepackage[dvipsnames]{xcolor}
\usepackage{lineno}
\usepackage[normalem]{ulem}

\arxivnumber{1234.56789} 
\title{Modelling Galactic neutrino emission: contributions from massive star clusters and interstellar cosmic rays}

\author[a,b]{S. Menchiari,}
\author[c,d]{S. Celli,}
\author[b,e]{V. Vecchiotti,}
\author[b]{G. Morlino,}
\author[b]{G. Peron}
\author[a]{and R. López-Coto}

\affiliation[a]{Instituto de Astrofísica de Andalucía, Glorieta de la Astronomía s/n, 18008, Granada, Spain}
\affiliation[b]{Istituto Nazionale di Astrofisica, Osservatorio Astrofisico di Arcetri, L.go E. Fermi 5, Firenze, Italy}
\affiliation[c]{Sapienza Universit\`a di Roma, Physics Department,
                 P.le Aldo Moro 5, 00185, Rome, Italy}
\affiliation[d]{ Istituto Nazionale di Fisica Nucleare, Sezione di Roma,
                 P.le Aldo Moro 5, 00185, Rome, Italy}
\affiliation[e]{Tsung-Dao Lee Institute, Shanghai Jiao Tong University, Shanghai 201210, P. R. China}

\emailAdd{smenchiari@iaa.es}
\emailAdd{silvia.celli@roma1.infn.it}

\abstract{The recent detection of Galactic neutrinos by the IceCube Observatory constitutes a remarkable achievement for neutrino astrophysics. By means of model dependent analyses based on spatial and spectral templates, built as to reproduce the GeV Galactic diffuse $\gamma$-ray data, a purely diffuse neutrino flux was measured in which no individual source was resolved. We present here a novel theoretical computation about the expected neutrino emission from the Galactic Plane that, differently from previous models, includes both the contributions from hadronic collisions of the cosmic-ray (CR) population and hadronic sources, represented by star clusters and supernova remnants therein, which are to date believed to be the dominant sources of Galactic CR protons. For the modelling of sources, diffusive particle acceleration is considered at both the collective wind termination shock blown by member stars and at the supernova shocks.  
The predicted flux of very-high energy neutrinos from individual star clusters is found to be marginally detectable even by cubic kilometer scale detectors, such that their cumulative contribution is expected to appear as an unresolved diffuse component, on top of that guaranteed by the CR sea interacting with the gas along the Plane. The overall neutrino production of the Milky Way star cluster population is computed, based on multiple synthetic realizations of the cluster population reproducing local stellar observations. As a result, we obtain novel neutrino template maps and provide them to the community, exploiting different assumptions with respect to the particle diffusion domain within clusters as well as the CR flux across the Galaxy, to be tested in future neutrino analyses in order to constrain the role of star clusters for extreme CR acceleration and neutrino production. The normalization of our models is consistent with the IceCube best-fit of existing Galactic templates, suggesting that the unresolved contribution from cluster emission may be non-negligible.  
}

\allowdisplaybreaks
\begin{document}
\maketitle
\flushbottom

\section{Introduction}
\label{sec:intro}
More than a century after the discovery of CRs, the origin of the most energetic particles known to date remains uncertain, limiting our understanding of 
the most efficient acceleration processes occurring in the Universe.
Several considerations indicate that charged particles up to energies of several petaelectronvolts (PeV) originate in supernova remnant (SNR) shocks, but $\gamma$-ray observations suggest that remnants of dead stars struggle to achieve PeV energies \citep{Cristofari2021}, pointing to the need for an alternative class of sources.
Recently, massive clusters of young stars (YMSCs) have been proposed as a viable complementary possibility in the search for PeVatrons \citep{aharonian2019}. From a theoretical perspective, compact clusters may offer more favourable environments for particle acceleration compared to isolated SNRs, due to the enhanced magnetic turbulence generated by colliding stellar winds and multiple SNR events at their cores \citep{vieu2022b}. The detection of non-thermal radiation from several YMSCs \citep{Peron_SCgamma_2025}, including the most energetic photon ever measured \citep{lhaasoCygnus}, suggests that efficient particle acceleration at PeV and beyond is occurring in these systems. At such extreme energies the radiation is expected to be most likely of hadronic origin: if confirmed, accelerated particles in these systems would be filling the transition gap between the knee and the ankle of the CR spectrum \citep{vieu2023}. \\
In hadronic scenarios, interactions between accelerated CRs and ambient gas produce charged and neutral pions, whose decays generate high-energy neutrinos and $\gamma$-rays, respectively. Differently from $\gamma$-rays, also radiated by leptons, high-energy neutrinos constitute an unambiguous probe of hadronic acceleration in astrophysical sources. In this context, the recent detection of high-energy ([1-100]~TeV) neutrinos from the Galactic Plane by the IceCube collaboration opens a new window for identifying hadronic particle accelerators in the Milky Way \citep{IceCubeScience}. This emission is expected to be contributed by both the diffuse CRs interacting along the Plane and in-situ production by - yet unresolved - hadronic accelerators. Disentangling the two components separately would provide key insights into the nature of CR sources in the Milky Way. 
In fact, the angular resolution offered by neutrino observations would help in probing individual CR sources, finally leading to the unequivocal identification of  Galactic accelerators. The weak signal provided by Galactic neutrinos, compared to the all-sky neutrino emission, has required dedicated analyses and more than 10 years of data taking to emerge. The methods adopted for its identification, so-called \emph{template fitting analyses}, strongly rely on a precise theoretical modelling for the spatial and spectral profile of the expected emission: so far, two different models have been tested, the so-called $\pi^0$ and KRA$_\gamma$ models, both accounting only for the purely diffuse neutrino emission generated by interaction of the CR sea with target gas in the Galactic Plane. The fitting procedure applied to the IceCube data has enabled the collaboration to extract the signal flux that is most compatible with the observations. Specifically, the different spatial and spectral templates adopted in the IceCube analyses lead to different event selections. While the fitted normalizations are comparable around $100$ TeV, the predictions diverge at lower energies because of the different spectral assumptions adopted in the two models.
However, because the models are tuned on the observed GeV $\gamma$-ray diffuse flux by Fermi-LAT and because the normalization of neutrino spectra is directly fitted to data, any unresolved source contributing to the flux is unavoidably hidden in the results.
Furthermore, independent searches for neutrino emission from known catalogued sources have not shown statistically significant excess of events from any  population in the Galaxy, which is possibly the result of the working hypothesis on the adopted templates \cite{IceCubeScience}. In order to overcome these limitations, we have developed a novel spectral and morphological template, embedding information about the emissivity of both the major CR candidate sources and the CR-induced neutrino diffuse component, that we strongly encourage to apply to neutrino datasets, aiming at precisely constraining the contribution of neutrino sources in the Galaxy as to finally probe CR sources. \\
Regarding the source contribution, because we are here concerned with the entire population of the Milky Way, we proceed to the realization of a synthetic cluster population, resembling the mass and radial distributions of the observed Galactic population, as described in Sec.~\ref{sec:SCsAndStars}. The overall sample of YMSCs is explored via a recently developed model for particle acceleration \citep{morlino2021}, describing both winds and SNRs therein. Because most of SN explosions occur in clusters, the latter are dominating the system energetics at later times, as discussed in Sec.~\ref{sec:cr}. Hadro-nuclear collisions of these accelerated particles with the target gas of the Milky Way are considered as the main radiative mechanism for  neutrino production \citep{mitchell2024}. Concerning the CR-induced neutrino emission, we exploit a range of different models, all being consistent with present CR data. The comparison among different spectral predictions of our model with IceCube current observation is provided in Sec.~\ref{sec:nu}, where the novel templates are also presented. 
Summary and conclusions about the role of unresolved neutrino sources are drawn in Sec.~\ref{sec:conclusions}. Additionally, in Appendix~\ref{app:B} we show the predicted templates as expected to be seen in large volume Cherenkov neutrino telescopes, under the different selection criteria of track-like and shower-like events, considering the angular resolution to these channels of IceCube and KM3NeT. Then, in Appendix~\ref{app:C} we discuss the impact of adopting a different gas map for the purely diffuse component. Finally, in Appendix~\ref{sec:appA} we provide predictions in terms of the $\gamma$-ray counterpart of the very same hadro-nuclear collisions producing neutrinos, and discuss the comparison with current diffuse $\gamma$-ray observations in the TeV-PeV energy range. 

\section{Simulation of a synthetic star clusters population}
\label{sec:SCsAndStars}

Motivated by the need to model the cumulative contribution of YMSCs to the Galactic neutrino emission, we construct a synthetic population of YMSCs representative of the Milky Way. Such an approach is required because the observed population of YMSCs is currently well characterized only in the local neighbourhood, within approximately 2 kpc from the Sun \citep{Piskunov2018, celli2024, Hunt2024}, preventing a direct, Galaxy-wide assessment of their contribution. To overcome this limitation, we generate 100 independent realizations of the Galactic YMSC population, sampling cluster and stellar properties according to observationally motivated distributions, as detailed in \cite{Menchiari25}, where the method has been first used for evaluating the Galactic hadronic $\gamma$-ray diffuse emission produced by the YMSC population. In the following, we briefly summarize the main steps of the method. 

\subsection{Generating star clusters}
\label{sec:SCspopulation}

As a first step in constructing the synthetic cluster population, we evaluate the total number of YMSCs in the Milky Way ($N_{\rm sc}$) as:
\begin{equation}
    N_{\rm sc}=\int_{M_{\min}}^{M_{\max}}\int_{0}^{t_{\rm max}}\int_{0}^{R_{\rm MW}} f(M_{\rm sc})\psi(t)\Sigma(r)\, dM_{\rm sc}\,dt\, dr 
\end{equation}
where $f(M_{\rm sc})$ is the cluster initial mass ($M_{\rm sc}$) function (IMF), $\Sigma(r)$ is the cluster formation rate as a function of the galactocentric radius $r$ and $\psi(t)$ is the cluster formation history as a function of time $t$.
We here employ the cluster IMF provided by \cite{Piskunov2018}, considering only clusters with masses ranging from $M_{\min }=1$~kM$_\odot$ to $M_{\max} =63$~kM$_\odot$, namely the maximum cluster mass observed in the Milky Way. The value of $M_{\min }=1$ is linked to the limited number of massive stars found in these objects, which results in very low wind power and a small number of supernova explosions. We consider only star clusters with age less than $t<t_{\max}=30$ Myr. The choice of $t_{\max}$ is related to the time needed by a 8 M$_\odot$ solar-metallicity star to exit the main sequence and explode as supernova \cite{Buzzoni2002}; in particular, clusters older than $t_{\max}$ are expected to contribute marginally to the total high-energy neutrino emission of the Galaxy, due to their lack of SNRs\footnote{In principle, supernova explosions can occur at later ages for various reasons, including stellar evolution in binary systems (which we do not treat here for simplicity) and the possibility that the star cluster formed over a prolonged period, rather than in a single starburst scenario where all stars are born at the same time.} and the weak power of stellar winds of their remaining  stars. $t_{max}$ is a relatively short timescale, when compared to the evolution timescale of  $\psi(t)$, and for this reason, the latter is approximated as a constant distribution \citep{Piskunov2018}, equal to current local formation rate \citep{Bonatto2011}. We consider instead the variation of $\psi$ as the Galactocentric radius, to scale as the radial distribution of giant molecular clouds.
To model this, we utilized the catalogue of giant molecular clouds provided by \cite{Hou2014} and averaged their masses across 18 distinct galactocentric rings. Each ring has a radial width of 1 kpc, extending up to a maximum Galactocentric radius of $R_{\rm MW}=18$ kpc.
 
The distribution is then normalized at the Sun distance so that it coincides with the cluster formation history inferred in the solar neighbourhood. As a result of these assumptions, we obtain $N_{\rm sc}= 2243$. 

\begin{figure}
    \centering
    \includegraphics[width=\textwidth]{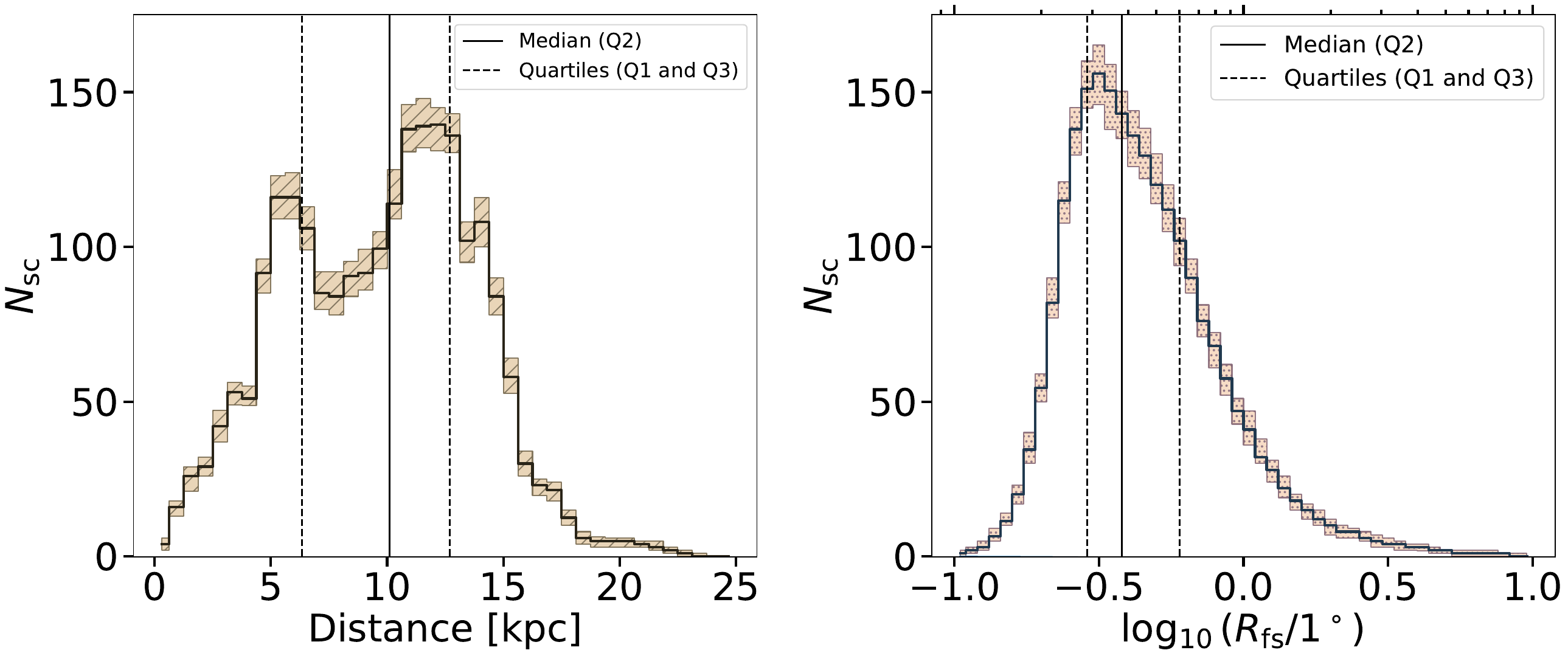}
    \caption{\emph{Left}: Distribution of heliocentric distances of the simulated star clusters. For each distance bin, we show the median (Q2) number of YMSCs computed over 100 independent realizations of the Galactic cluster population, as well as the 25th–75th percentile range (Q1 and Q3) plotted as a shaded region. \emph{Right:} Distribution of the projected bubble radius on the sky for the simulated SC population, with the shaded region indicating the 25th–75th percentile range (Q1 and Q3) as computed over 100 independent realizations of the Galactic population. The vertical black solid and dashed lines mark the median (Q2) and the two quartiles of the overall distribution, respectively. The size of the bubbles have been calculated including cooling losses by rescaling the total mechanical luminosity $L_{\rm tot}$ with a constant factor $\eta_{\rm{m}}=0.1$, see text for further details.}
    \label{fig:Dist_Size_Distrib}
\end{figure}

Once the total number of relevant YMSCs is known, we proceed to the simulation of 100 distinct populations, by extracting the mass and age of each object by random sampling $f(M_{\rm sc})$ and $\psi(t)$ respectively. The position of each YMSC in the Milky Way is instead obtained using a more refined approach: firstly, the galactocentric radial and angular coordinate along the plane are extracted using $\Sigma(r)$ while assuming an isotropic angular distribution; subsequentially, we use these coordinates to associate each YMSC with a specific Galactic structure. We here consider a model of the Milky Way that includes the following structures: the four main spiral arms \citep{Hou2014}, the Local Spur \citep{Hou2014}, the Galactic bar \citep{Churchwell2009}, and inner arm segments \citep{Churchwell2009}. The resulting distribution in the Galactic Plane of the simulated YMSCs is provided in the left panel of Fig.~\ref{fig:Dist_Size_Distrib}, where their heliocentric distance is shown with a shaded band representing the resulting uncertainty from the 100 realizations.

\subsection{Simulating stars and their winds}
For every synthetic YMSC, a mock population of stars is produced by random sampling the stellar IMF $f_\star(M_\star)$ \citep{Kroupa2001} in the stellar mass range $M_\star \in [M_{\star,\min}, M_{\star,\max}]$, where $M_{\star,\min}=0.08$~M$_\odot$ is the minimum mass required for hydrogen burning \citep{Carroll_IntroModAstro_1996}, while $M_{\star,\max}=150$~M$_\odot$ is the maximum stellar mass observed in the Milky Way \citep{Zinnecker2007}. The number of stars inside each cluster is then calculated as:
\begin{equation}
    N_\star=  M_{\rm sc} \frac{\int_{M_{\star,\min}}^{M_{\star,\max}} f_\star(M_\star)dM_\star}{\int_{M_{\star,\min}}^{M_{\star,\max}} M_\star f_\star (M_\star) dM_\star} \, .
\end{equation}
Once the stellar population is extracted, we evolve it accordingly to the age of the synthetic cluster ($t_{\rm sc}$). To do so, we calculate for each mock star the main sequence turn off time ($\tau_{\rm ms}$) \citep{Buzzoni2002}, defined as the age at which a star leaves the main sequence (MS):
\begin{equation}
\log_{10} \left (\frac{\tau_{\rm ms}}{1\, \rm yr} \right ) = 0.825 \log_{10}^2 \left (\frac{M_\star}{120\, \rm M_\odot} \right ) + 6.43 \,,
\end{equation}
and, by using this parameter, we classify the stars into three categories:
\begin{enumerate}
    \item MS stars: all stars with $t_{\rm sc}<\tau_{\rm ms}$ are considered to be still in the MS.
    \item Wolf-Rayet (WR) stars: stars with $t_{\rm sc}-\tau_{\rm ms}<0.3$~Myr and $M_\star>25$~M$_\odot$ have left the MS, and are considered to be in their WR phase. Note that we do not include any post MS evolutionary stage for lower-mass stars.
    \item Supernovae (SNe): all stars with mass $>8 M_{\odot}$ that do not fall in the previous categories are considered to have exploded as SNe. 
\end{enumerate}

According to the evolutionary stage of each star (MS or WR), we calculate the mass loss rate ($\dot{M}_{\star}$) of stars using a purely empirical approach \citep[see][and references therein]{Menchiari25}. For the wind speed ($v_{\rm w,\star}$) of MS stars, we use the prescription reported in \cite{Kudritzki2000}, while an average constant value of 1500~km~s$^{-1}$ is adopted for WR stars. The wind power of each star is then computed as $L_{\rm w,\star}=\frac{1}{2}\dot{M}_{\star}v_{\rm w,\star}^2$. Finally, we calculate the total wind luminosity ($L_{\rm w}$), mass loss rate ($\dot{M}$) and collective wind speed ($v_{\rm w}$) of a given YMSC as:
\begin{equation}
\label{eq:vwind}
\begin{cases}
        L_{\rm w}=\sum_{i=0}^{N_\star}(L_{\rm w,\star})_{i} \\
        \dot{M}=\sum_{i=0}^{N_\star}(\dot{M}_{\star})_{i}
\end{cases}
\implies         v_{\rm w}=\sqrt{\frac{2 \dot{M}}{L_w}}\, .
\end{equation}
We further account for the power injected by SNe in the system by calculating their average luminosity per cluster as
\begin{equation}
    L_{\rm sn}=\frac{E_{\rm sn} N_{\rm sn}}{t_{\rm sc}}
\end{equation}
where $E_{\rm sn}=10^{51}$~erg is the kinetic energy released by each SN explosion and $N_{\rm sn}$ is the total number of SNe exploded in the YMSC. 
 
\begin{figure}
    \centering
    \includegraphics[width=\textwidth]{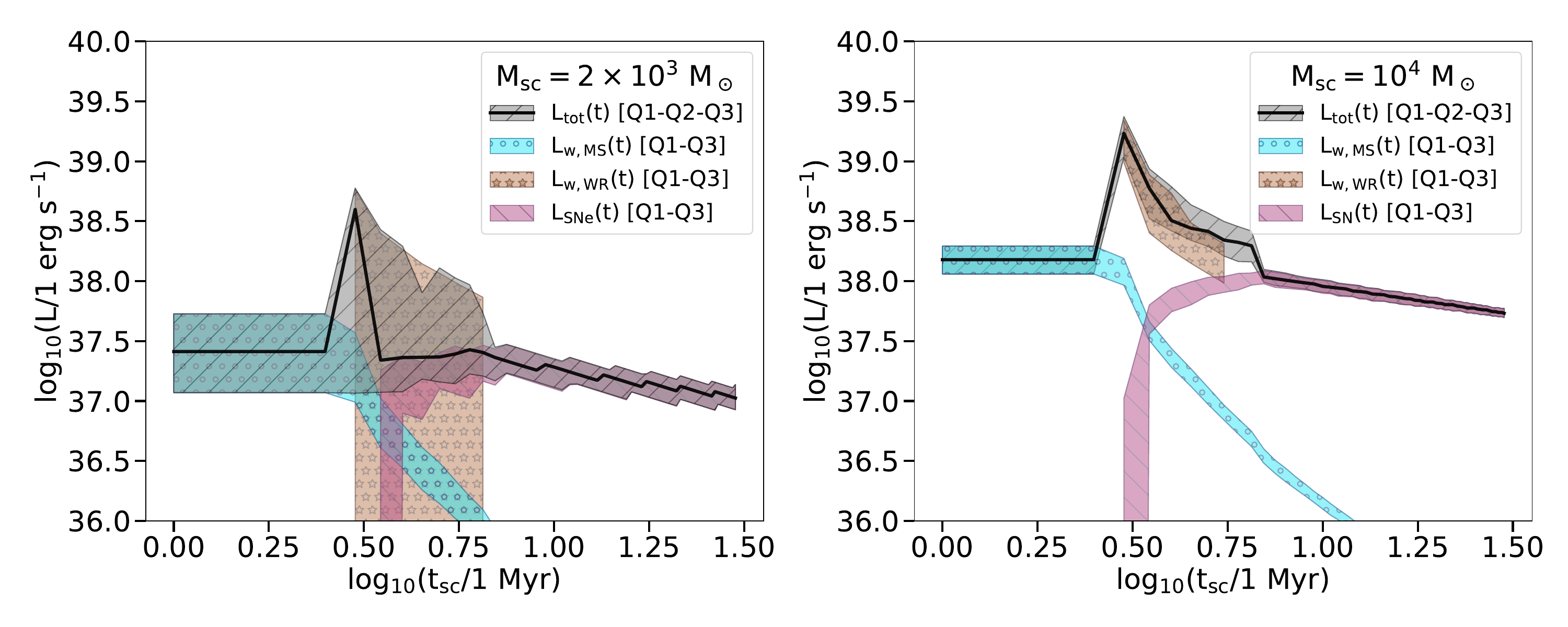}
    \caption{Time evolution of the mechanical power injected by a star cluster as a function of its age. The cyan band with dotted hatching shows the mechanical power from MS stellar winds ($L_{\rm w, \, MS}$), the brown band with star hatching corresponds to winds from WR stars ($L_{\rm w, \, wr}$), and the purple band with backslash hatching represents the average mechanical power by SNe ($L_{\rm SNe}$). The black band with forward-slash hatching shows the total mechanical power ($L_{\rm tot}$), obtained as the sum of all contributions. The solid black line indicates the median of the total power (Q2). The shaded bands enclose the 25th–75th percentile range (Q1-Q3), derived from 500 random samplings of the stellar initial mass function. The left and right panels refer to a cluster mass of $2 \times 10^3$ M$_\odot$ and $1 \times 10^4$ M$_\odot$, respectively.} 
    \label{fig:MSWRSNpower}
\end{figure}

The resulting cluster luminosity as a function of time is presented in Fig.~\ref{fig:MSWRSNpower}, where the individual contributions of MS and WR stars is shown, as well as that of SNe, for two representative values of cluster masses, namely $2 \times 10^3$ M$_\odot$ and $1 \times 10^4$ M$_\odot$.

\subsection{Modelling bubbles around star clusters}
\begin{figure}
    \centering
    \includegraphics[width=\textwidth]{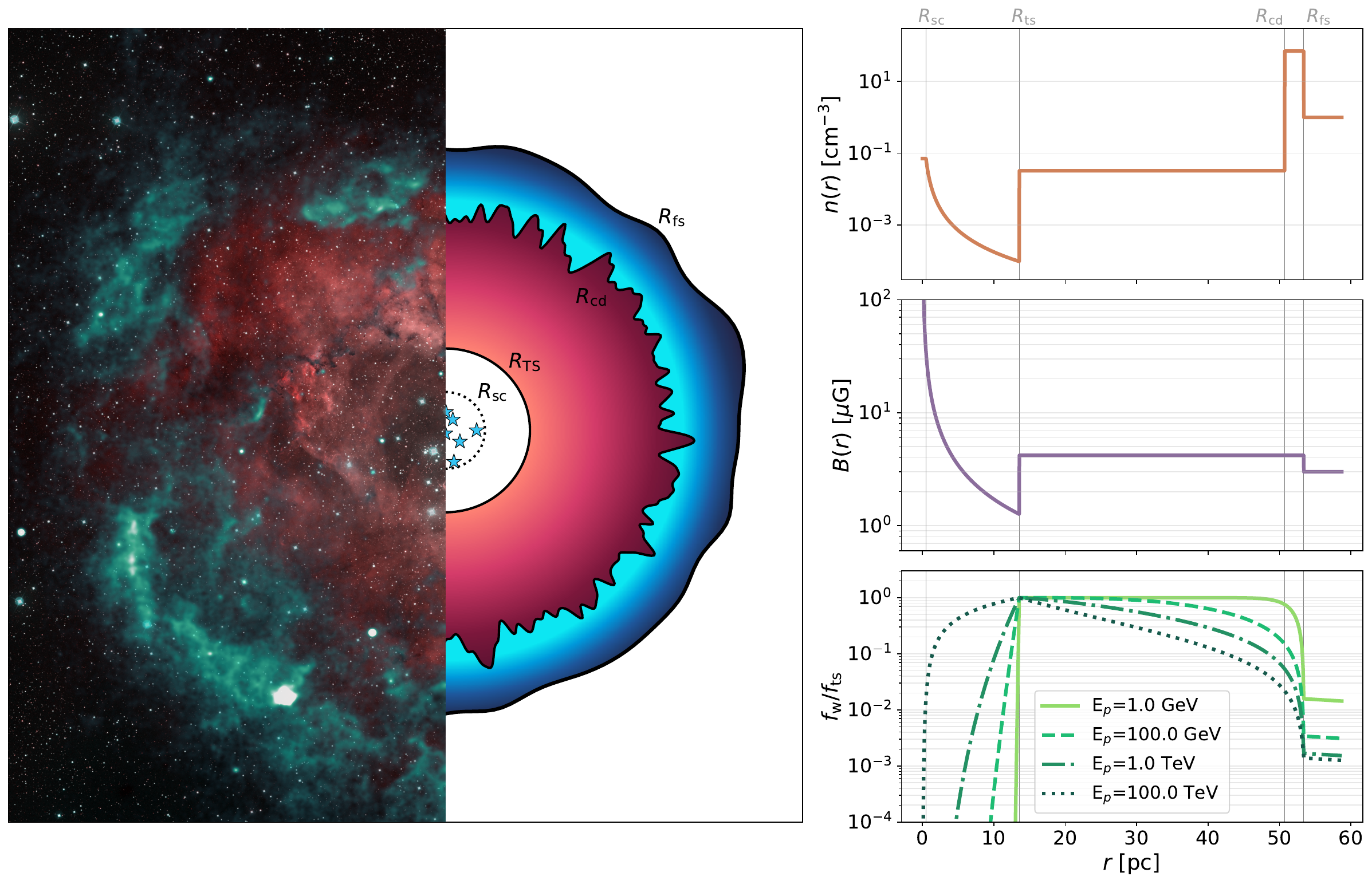}
    \caption{\emph{Left:} False-color image of the Rosette nebula: red is optical (DSS2 sky survey) and green/lightblue is infrared (MSX band A), next to a schematic representation of the wind-blown bubble. \emph{Right:} Radial profiles of gas number density, magnetic field strength and accelerated particle density (normalized to its value at the WTS) at different energies
    for a benchmark cluster with $L_w = L_{sn}=10^{37}$ erg s$^{-1}$, $\dot{M}=10^{-5}$ M$_\odot$ yr$^{-1}$ and $t_{sc}=5$ Myr. }
    \label{fig:sketch}
\end{figure}

The combined feedback of stellar winds and supernova explosions creates large bubbles in the interstellar medium (ISM), which are generally referred to as wind-blown bubbles if they are mostly powered by stellar winds, or superbubbles if the energy budget is provided by both SNe and winds. In compact star clusters\footnote{We here refer to  ``compact'' star clusters as all those clusters where the winds from their massive stars merge to form a collective outflow. Although there is no general consensus, a possible criterion to quantify the occurrence of this condition is that the cluster core radius must be smaller than the wind termination shock radius. In this work, we will assume that all clusters satisfy this criterion.}, massive star winds merge into a supersonic collective outflow, producing bubbles with dynamics and structure similar to those around isolated massive stars \citep{Weaver1977}. The structure of these objects is divided in four distinct regions (see Fig.~\ref{fig:sketch}): (i) the innermost core, with size $R_{\rm sc}$, where stars reside and individual stellar winds are launched; (ii) the free–flowing collective cluster wind, confined by the wind termination shock (WTS) located at $R_{\rm ts}$; (iii) a cavity composed by hot shocked plasma, filling the bubble interior between the WTS and the contact discontinuity at $R_{\rm cd}$; and (iv) a dense shell of swept–up interstellar material created by the expansion of the hot cavity, bounded internally by the contact discontinuity and externally by the forward shock at $R_{\rm fs}$, which delimits the overall structure from the ambient ISM. The location of the WTS is set by the balance between the ram pressure of the collective cluster wind and the thermal pressure of the shocked hot gas, and can be expressed as:
\begin{equation}
    R_{\rm ts} =\sqrt{\mathcal{A}}\; (\dot M v_{\rm w})^{1/2}\; (\eta_{\rm m} L_{\rm tot})^{-1/5} \; \rho_0^{-3/10} \; t_{\rm sc}^{2/5} ,
\end{equation}
where $\mathcal{A}= \frac{(3850\pi)^{2/5}}{28 \pi}$ and $\rho_0$ is the mass density of the environment in which the bubble expands; while the position of the forward shock and the contact discontinuity are respectively:
\begin{equation}
    R_{\rm fs} =  \left (\frac{125}{154 \pi} \right )^{1/5} \left ( \eta_{\rm m} L_{\rm tot} \right )^{1/5} \rho_0^{-1/5} t_{\rm sc}^{3/5}
\end{equation}
and $R_{\rm cd} \simeq 0.95 \, R_{\rm fs}$. We note that the former expressions of $R_{\rm ts}$ and $R_{\rm fs}$ differ from the ones obtained for a stellar wind-blown bubble as derived in \cite{Weaver1977} by the presence of the parameter $\eta_{\rm m}$: this represents the mechanical efficiency, namely the fraction of the wind kinetic energy effectively used to blow the bubble, i.e. $L_{\rm kin} = \eta_{\rm m} L_{\rm tot}$, where $L_{\rm tot}=L_{\rm sn}+L_{\rm w}$. The remaining fraction $(1-\eta_{\rm m}L_{\rm tot})$ is lost due to radiative cooling. Numerical simulations suggest $\eta_{\rm m}$ of the order of a few tens of percent \citep{Vasiliev2015, Yadav2017}, and in our calculations we adopt a fixed value of $\eta_{\rm m}=0.1$.
The effect of $\eta_{\rm m}$ is to reduce the bubble pressure to the following value \citep{morlino2021}:
\begin{equation}
    P_b= \frac{7}{(3850 \pi)^{2/5}} (\eta_{\rm m} L_{\rm tot})^{2/5}\rho_0^{3/5}t_{\rm sc}^{-4/5} \, .
\end{equation}
The distribution of bubble radii for the synthetic YMSC samples is shown in Fig.~\ref{fig:Dist_Size_Distrib}: it can be noted that, being the simulated population spread over the entire Galaxy, their angular sizes are noticeably smaller than, e.g. the nearby Gaia sample \cite{mitchell2024}.

Particularly relevant for hadronic emission is the distribution of target material that regulates the probability of proton-proton collisions. The density profile in wind-blown bubbles is \citep{Weaver1977}:
\begin{equation}
n(r) =
\begin{cases}
 \dot{M}\Big/(4\pi R_{\rm sc}^2 v_{\rm w} m_{\rm p})               & r < R_{\rm sc} \\[0.5em]
\dot{M}\Big/(4\pi r^2 v_{\rm w} m_{\rm p})                        & R_{\rm sc} \leq r < R_{\rm ts}\\[0.5em]
M_{\rm c} \Big/ \left[ \frac{4}{3}\pi (R_{\rm cd}^3-R_{\rm ts}^3)m_{\rm p} \right]        & R_{\rm ts} \leq r < R_{\rm cd}\\[0.5em]
 n_0 \Big/ \left ( 1-\frac{R_{\rm cd}^3}{R_{\rm fs}^3} \right) - \dot{M}_{\rm sh} t_{\rm sc} \Big/\left[ \frac{4}{3}\pi (R_{\rm fs}^3-R_{\rm cd}^3)m_{\rm p} \right]                                         & R_{\rm cd} \leq r \leq R_{\rm fs}~,
\end{cases}
\end{equation}
where $m_p$ is the proton mass, $M_{\rm c}=(\dot{M}+\dot{M}_{\rm sh}) t_{\rm sc}$ is the total mass enclosed in the hot cavity, $v_{\rm{w}}$ is the cluster wind velocity as defined in Eq.~\eqref{eq:vwind} and $\dot{M}_{\rm sh}$ is the mass evaporation rate of the shell \citep{Castor1975}. Most of the target material is enclosed in the dense shell, and the total swept-up mass is directly related to the ambient particle density that we here consider to be $n_0=10$~cm$^{-3}$ for all YMSCs. Finally, we note that the possible fragmentation of the swept-up shell caused by the onset of hydrodynamical instabilities might raise $M_{\rm c}$, effectively incrementing the density in the cavity. We do not account for this effect here; in this sense, the density we adopt for the cavity should be considered as a lower limit. 

Of similar importance is the radial profile of the magnetic field strength, which characterizes the normalization of the particle diffusion coefficient in the system. We here consider that the magnetic field is injected at the centre of the cluster and advected outward by the collective wind, leading to a radial dependence proportional to $1/r$ in the upstream region \citep{morlino2021}. At the WTS, the magnetic field is compressed by a factor $\sqrt{11}$, and in the downstream it fills the bubble with such a uniform intensity. The absolute normalization is obtained by assuming that the magnetic pressure downstream of the WTS is a fraction $\eta_{\rm B}$ of the total bubble pressure, i.e:

\begin{equation}
    \frac{B^2}{8 \pi}= \eta_{\rm B} P_b \, .
\end{equation}

The resulting profile is then: 
\begin{equation}
B(r) =
\begin{cases}
\sqrt{\mathcal{A}'} \, \eta_{\rm B}^{1/2}(\eta_{\rm m} L_{\rm tot})^{1/5}\rho_0^{3/10}t_{\rm sc}^{-2/5} \left (\frac{R_{\rm ts}}{r} \right )             & r \leq R_{\rm ts}\\[0.5em]
\left [\frac{56 \pi}{(3850 \pi)^{2/5}}\right ]^{1/2}\eta_{\rm B}^{1/2}(\eta_{\rm m} L_{\rm tot})^{1/5}\rho_0^{3/10}t_{\rm sc}^{-2/5}                      & R_{\rm ts} < r < R_{\rm fs},
\end{cases}
\end{equation}
where we have defined $\mathcal{A}' \equiv \frac{56 \pi}{11(3850 \pi)^{2/5}}$. We note that the normalization of the magnetic field adopted in this work differs from that used in \cite{Menchiari25}. Here, the magnetic field is normalized to the plasma pressure downstream of the WTS, a choice motivated by the need to account for the energy input from SNe, which was not included in \cite{Menchiari25}, while there the magnetic field strength was obtained by converting a fraction $\eta_B'$ of the wind kinetic luminosity in magnetic luminosity. The two approaches give the same result when there are no SNe by imposing $\eta_B = \frac{\eta_B'}{2 \mathcal{A} \mathcal{A}'} = 11/4 \, \eta_B'$. In this work, we consider a value of $\eta_B = 0.14$, corresponding to $\eta_B'\simeq 0.05$.
For illustrative purposes, we show in Fig.~\ref{fig:sketch} the radial profiles of gas number density, magnetic field strengths and CR distribution function, for a reference cluster with $L_w = L_{sn}=10^{37}$ erg s$^{-1}$, $\dot{M}=10^{-5}$ M$_\odot$ yr$^{-1}$ and $t_{sc}=5$ Myr. The calculation of the resulting CR distribution is described in the following Section.

\section{CR production in star clusters}
\label{sec:cr}
Particle acceleration in YMSCs can be powered by the strong stellar winds and SNe exploding in the core. Therefore in the following when referring to YMSCs, we will be referring to the summed contribution, unless specifically stated. The relative contribution of these two engines is not constant in time: stellar winds dominate CR production in the first few Myr, but as massive stars evolve and SNe begin to occur, the wind power declines, eventually leading SNe to become the dominant source of acceleration \citep{Vieu2022a}, as we show in Fig.~\ref{fig:MSWRSNpower}. We model CR production and propagation within the cluster bubble by combining the theoretical framework of \cite{morlino2021} for stellar wind acceleration and transport, with the approach of \cite{mitchell2024} where the average SN contribution was considered as an extra source of power to the wind component. We summarize below both models.

\subsection{Particle acceleration by stellar winds}
In compact YMSCs, particle acceleration may occur at the collective cluster WTS. This scenario has been described in a recent work by \cite{morlino2021}, where the authors solved the particle transport  under the assumptions of steady-state injection of protons, radial symmetry, and negligible energy losses, apart from adiabatic ones. Therefore, all of the following discussion holds for protons. The resulting CR radial distribution in the system is:
\begin{equation}
\label{eq:fcr}
f_{\rm w}(r, p) =
\begin{cases}
f_{\rm ts}(p) \cdot \exp \left [ -\int_r^{R_{\rm ts}} \frac{v_{\rm w}}{D(r',p)}  dr' \right ]   & r < R_{\rm ts} \\[0.75em]
f_{\rm ts}(p) \frac{ \left[e^{\alpha} + \beta(e^{\alpha_{\rm b}} - e^{\alpha}) \right]}{1+\beta(e^{\alpha_{\rm b}}-1)}  & R_{\rm ts} \leq r \leq R_{\rm b}~,
\end{cases}
\end{equation}
where $f_{\rm ts}(p)$ represents the accelerated particle spectrum at the WTS location as a function of particle momentum $p$, while the functions $\alpha$, $\alpha_{\rm b}$ and $\beta$ are defined as:
\begin{subequations}
\label{eq:AlphaBeta}
\begin{equation}
\alpha \equiv \alpha(r, p)=\frac{v_{\rm b} R_{\rm ts}}{D(r,p)} \left ( 1 -  \frac{R_{\rm ts}}{r} \right )
,\end{equation}
\begin{equation}
\alpha_{\rm b}\equiv\alpha(r=R_{\rm b},p)
,\end{equation}
\begin{equation}
\beta \equiv \beta(p) = \frac{D_{\rm ism}(p) R_{\rm b}}{v_{\rm b} R_{\rm ts}^2} \,.
\end{equation}
\end{subequations}
with $v_{\rm b}=v_{\rm w}/4$ as the plasma speed downstream the WTS. In Eqs.~\eqref{eq:AlphaBeta} and following, the parameters $D_{\rm ism}(p)=3\times10^{28}(pc/1 \rm{~GeV})^{1/3}$~cm$^2$s$^{-1}$ and $D(r,p)$ represent the particle diffusion coefficient in the ISM and inside the bubble, respectively. The spectral shape of the latter depends on the type of turbulence spectrum developing in the bubble, which is still unknown. Therefore, in this work, we consider three distinct scenarios related to a Kolmogorov, Kraichnan, and Bohm turbulence, each yielding a different diffusion coefficient, parametrized as follows:
\begin{subequations}
\label{eq:DiffCoeff}
\begin{equation}
D_{\rm Kol}(r, p)=\frac{1}{3} v_{\rm p} r_{\rm L}(r, p)^{1/3} L_{\rm inj}^{2/3}
,\end{equation}
\begin{equation}
D_{\rm Kra}(r, p)=\frac{1}{3} v_{\rm p} r_{\rm L}(r, p)^{1/2} L_{\rm inj}^{1/2}
,\end{equation}
\begin{equation}
D_{\rm Bohm}(r, p)=\frac{1}{3} v_{\rm p} r_{\rm L}(r, p)  \,.
\end{equation}
\end{subequations}
where $v_{\rm p}\simeq c$ is the particle velocity and $c$ is the in-vacuum speed of light, $L_{\rm inj}$ is the injection length scale of the turbulence\footnote{We assume that the turbulence is injected for all clusters at a typical length scale of 1~pc, corresponding approximatively to the mean size of the cluster core (or to the average distance between stars) observed in open clusters \citep{Tarricq2022, Just2023}.} and $r_{\rm L}=\frac{pc}{q_e B(r)}$ is the particle Larmor radius with $q_e$ being the electron charge. From this point forward, all results are presented for three distinct cases, each corresponding to one of the assumptions regarding the diffusion coefficient.

The spectrum of injected particles at the WTS, $f_{\rm ts}$, results from the numerical solution of the particle transport equation \citep{morlino2021}, and it depends on the nature of particle diffusion coefficient. In this work, we adopt an analytical approximation obtained by fitting the formal numerical solution, as described in \cite{Menchiari2024}, namely:
\begin{equation}
    f_{\rm ts}(p) = \frac{3 n_{\rm w} v_{\rm w}^2 \varepsilon_{\rm cr, w}}{4 \pi \Lambda_{\rm w} (m_{\rm{p}}c)^3 c^2 } \left( \frac{p}{m_{\rm{p}} c} \right)^{-s_{\rm w}} \, \left [ 1 + a_1 \left (\frac{p}{p_{\max,\rm{w}}} \right )^{a_2} \right ] \mathrm{e}^{- a_3 (p/p_{\max,\rm{w}})^{a_4}} \,,
    \label{eq:fts}
\end{equation}
where $n_{\rm w}=\frac{\dot{M}}{4 \pi R_{\rm ts}^2v_{\rm w}}$ is the number density immediately upstream of the WTS, the coefficients $a_{1, ...,4}$ are the ones that provide the best fit to the formal equation \cite[see Table~1 in][]{Menchiari2024}, and the scalar $\Lambda_w$ is the defined as:
\begin{equation} \label{eq:Lambda}
 \Lambda_{\rm w} =\int_{x_{\rm inj}}^{\infty} \frac{x^{4-s_{\rm w}}}{\sqrt{1+x^{2}}} \left [ 1 + a_1 \left (\frac{x}{x_{\max,\rm{w}}} \right )^{a_2} \right ] \mathrm{e}^{- a_3 (x/x_{\max,\rm{w}})^{a_4}} dx \,,
\end{equation}
with $x=p/(m_p c)$, $x_{\rm inj}\equiv x(pc=1 \rm{~GeV})$ and $x_{\max,\rm{w}}\equiv x(p=p_{\max,\rm{w}})$. Finally, the parameters $\varepsilon_{\rm cr, w}$, $s_{\rm w}$ and $p_{\max,\rm{w}}$ are respectively the CR acceleration efficiency\footnote{We here define the efficiency of particle acceleration as the fraction of wind momentum flux converted into CR pressure.}, the spectral slope of injected particles and the maximum momentum produced at the WTS. The acceleration efficiency of stellar winds estimated from $\gamma$-ray observations ranges from a few percent to ten percent \citep{aharonian2019, Peron2024, Peron2025}, while the injection spectrum is well traced by the slope of the non-thermal emission at high energies, which is measured to range from $-4$ to $-4.4$ \citep{Yang2018, Yang2020, Saha2020, Astiasarain2023, Peron2024, Peron2025}. 
We hence consider for all YMSCs an acceleration efficiency of $\varepsilon_{\rm cr, w}=0.1$ and a spectral slope of $s_{\rm w}=4.2$. The maximum particle momentum depends on the confinement of CRs upstream the WTS, and it is settled by the condition $D(r=R_{\rm ts},p_{\max,\rm{w}})/v_{\rm w} = R_{\rm ts}$. Given the definition of the diffusion coefficients in Eqs.~\eqref{eq:DiffCoeff}, the maximum momenta in the three different turbulence scenarios are:
\begin{subequations}
\begin{equation}
p_{\max,\, \mathrm{Kol}} =
\left[ \frac{27 \, q_e \, \mathcal{A}^{3/2} \sqrt{\mathcal{A}'}}{c^4}  \right] \eta_{\rm B}^{1/2}
\; \dot M^{3/2} v_{\rm w}^{9/2} \;
(\eta_{\rm M} L_{\rm tot})^{-2/5} \; \rho_0^{-3/5} \; t_{\rm sc}^{4/5} L_{\rm inj}^{-2}
,\end{equation}
\begin{equation}
 p_{\max,\, \mathrm{Kra}} =
\left[ \frac{9 \, q_e \, \mathcal{A} \sqrt{\mathcal{A}'}}{c^3} \right] \eta_{\rm B}^{1/2}
\; \dot M \, v_{\rm w}^3 \;
(\eta_{\rm M} L_{\rm tot})^{-1/5} \; \rho_0^{-3/10} \; t_{\rm sc}^{2/5} L_{\rm inj}^{-1}
,\end{equation}
\begin{equation}
p_{\max, \, \mathrm{Bohm}} =
\left[ \frac{3 \, q_e \, \sqrt{\mathcal{A} \mathcal{A}'}}{c^2} \right ] \eta_{\rm B}^{1/2} \; \dot M^{1/2} \, v_{\rm w}^{3/2}
\end{equation}
\end{subequations}
which can be rewritten in a more convenient form as:
\begin{subequations}
\begin{align}
\begin{aligned}
p_{\max,\, \mathrm{Kol}} \simeq 
0.45 \,
\left(\frac{\eta_B}{0.14}\right)^{1/2}
\left(\frac{\dot{M}}{10^{-5}\,M_\odot\,{\rm yr}^{-1}}\right)^{3/2}
\left(\frac{v}{10^3\,{\rm km\,s^{-1}}}\right)^{9/2} \times \\
\times
\left(\frac{\eta_M L_{\rm tot}}{0.1 \times 10^{37}\,{\rm erg\,s^{-1}}}\right)^{-2/5}
\left(\frac{\rho_0}{10\,m_p\,{\rm cm^{-3}}}\right)^{-3/5}
\left(\frac{t_{\rm age}}{1\,{\rm Myr}}\right)^{4/5}
\left(\frac{L_{\rm inj}}{1\,{\rm pc}}\right)^{-2}
\; {\rm TeV/c}
\end{aligned}
\\[1em]
\begin{aligned}
p_{\max,\, \mathrm{Kra}} \simeq 
7.4 \,
\left(\frac{\eta_B}{0.14}\right)^{1/2}
\left(\frac{\dot{M}}{10^{-5}\,M_\odot\,{\rm yr}^{-1}}\right)
\left(\frac{v}{10^3\,{\rm km\,s^{-1}}}\right)^{3} \times \\
\times
\left(\frac{\eta_M L_{\rm tot}}{0.1 \times 10^{37}\,{\rm erg\,s^{-1}}}\right)^{-1/5}
\left(\frac{\rho_0}{10\,m_p\,{\rm cm^{-3}}}\right)^{-3/10}
\left(\frac{t_{\rm age}}{1\,{\rm Myr}}\right)^{2/5}
\left(\frac{L_{\rm inj}}{1\,{\rm pc}}\right)^{-1}
\; {\rm TeV/c}
\end{aligned}
\\[1em]
\begin{aligned}
p_{\max, \, \mathrm{Bohm}} \simeq 
120 \,
\left(\frac{\eta_B}{0.14}\right)^{1/2}
\left(\frac{\dot{M}}{10^{-5}\,M_\odot\,{\rm yr}^{-1}}\right)^{1/2}
\left(\frac{v}{10^3\,{\rm km\,s^{-1}}}\right)^{3/2}
\; {\rm TeV/c}
\end{aligned}
\end{align}
\end{subequations}

\subsection{Particle acceleration at supernovae remnant shocks}
After $\sim3$~Myr the most massive stars end their post-MS evolution and explode as SNe \citep{Buzzoni2002}. The modelling of CR acceleration by SNRs in wind-blown bubbles is complicated by  the non-uniform structure of the cluster environment, as well as the transient nature of SNR evolution, which would require a full time-dependent treatment. Nevertheless, an approximate estimation of their contribution to particle acceleration can be obtained by treating SNRs as a continuous, time-averaged sources of CRs, following the same approach reported in \cite{mitchell2024}. 

SNRs can accelerate CRs mainly during both the ejecta-dominated (ED) and the Sedov-Taylor (ST) phases\footnote{The ED phase refers to the early stage of a SNR evolution, also known as free expansion phase, when the ejecta mass exceeds the swept-up ambient mass. The ST phase occurs immediately after the ED phase, when the swept-up mass becomes larger than the ejecta's and the shock evolves adiabatically.}, however the duration of the latter phase is reduced with respect to SNR evolving in the typical ISM, due to the larger bubble temperature. As a consequence the bulk of CRs is accelerated during the ED phase, where the shock speed is roughly constant and the CR spectrum can be approximated as:
\begin{equation}    
\label{eq:f_snr}
    f_{\rm snr}(p) = \frac{3 \, \varepsilon_{\rm cr, sn} n_{\rm b} u_{\rm ed}^2}{4\pi\, \Lambda_{\rm{sn}} (m_{\rm{p}} c)^4 c^2} \left( \frac{p}{m_{\rm{p}} c}\right)^{-s_{\rm sn}} \mathrm{e}^{-p/p_{\max,\rm{sn}}} \,,
\end{equation}
where $n_{\rm b}\equiv n(R_{\rm ts}\leq r<R_{\rm cd})$ is the density downstream the WTS and $u_{\rm ed}=(2 E_{\rm sn}/M_{\rm ej})^{1/2}$ is the shock speed in the ED phase ($M_{\rm ej}$ being the ejecta mass\footnote{The amount of ejecta mass in a SN explosion is highly debated, mainly due to the uncertainty in the mass loss rate during the final phase of the massive stellar evolution. For the sake of simplicity, we will assume that all SNe produce an ejecta mass of 5~M$_\odot$ \cite[see, e.g.][]{Limongi-Chieffi:2010}.}). The parameters $\varepsilon_{\rm cr, sn}=0.1$ and $s_{\rm sn}=4.3$ are respectively the CR acceleration efficiency and the spectral injection slope, whose values are chosen following the properties of observed SNRs \citep{Funk2015}. The scalar $\Lambda_{\rm{sn}}$ in Eq.~\eqref{eq:f_snr} is:
\begin{equation} \label{eq:Lambda}
 \Lambda_{\rm sn} =\int_{x_{\mathrm{inj}}}^{\infty} 
\frac{x^{4-s_{\rm sn}}}{\sqrt{1+x^{2}}} \, 
e^{-x/x_{\max, sn}} \, dx  \,,
\end{equation}
with $x_{\max,\rm{sn}}\equiv x(p=p_{\max,\rm{sn}})$. The parameter $p_{\max,\rm{sn}}$ represents the maximum momentum of particles accelerated at SNRs, which is fixed by a time-limited condition, i.e. by equating the acceleration time to the start of the ST phase. The acceleration time is directly related to the magnetic field downstream the WTS, where most of ED phase of SNRs occurs. However, on top of that, magnetic field amplification can develop due to streaming instabilities, self-induced by the very same accelerated particles. Considering only resonant streaming instabilities\footnote{Non-resonant streaming instabilities might also play a role. However, they dominate the amplification of magnetic fields under extreme conditions of very fast shocks and high upstream densities, which may occur only during the first hundreds years of evolution \citep{Bell2013}, a phase that we neglect in the present calculation.}, the maximum particle momentum reads as:  
\begin{equation}
\begin{aligned}
p_{\max,\rm{sn}}\simeq290~
    \left( \frac{\varepsilon_{\rm cr, sn}}{0.1} \right)^{1/2}
    \left( \frac{E_{\rm sn}}{10^{51} \rm erg} \right)^{1/4}
    \left( \frac{M_{\rm ej}}{5\, \rm{M}_{\odot}} \right)^{-1/4} \times \\
    \times \left( \frac{B_{\rm st}}{10 \, \rm \mu G} \right) 
    \left( \frac{u_{\rm ed}}{5000 \, \rm km\, s^{-1} } \right)
    \left( \frac{R_{\rm st}}{10 \, \rm pc} \right) \, \rm TeV/c,   
\end{aligned}
\end{equation}
where $R_{\rm st}$ is the size of the SNR when it enters the ST phase, namely:
\begin{equation}
R_{\rm st} = 
\begin{cases}
R_{\rm ts} + \left[\frac{3}{4 \pi n_{\rm b} m_{\rm p}} \left( M_{\rm ej} - \frac{\dot{M} R_{\rm ts}}{v_{\rm w}} \right) \right]^{1/3}               & \text{if: }  M_{\rm ej} \geq \frac{\dot{M} R_{\rm ts}}{v_{\rm w}}\\[0.5em]
v_{\rm w }M_{\rm ej} \Big / \dot{M}  & \text{if: } M_{\rm ej} < \frac{\dot{M} R_{\rm ts}}{v_{\rm w}}\, ,
\end{cases}
\end{equation}
and $B_{\rm st}$ is the magnetic field strength upstream the SNR at the moment that it enters the ST phase, that is:
\begin{equation}
B_{\rm st} = 
\begin{cases}
B(R_{\rm ts}<r<R_{\rm cd})               & \text{if: }  M_{\rm ej} \geq \frac{\dot{M} R_{\rm ts}}{v_{\rm w}}\\[0.5em]
B(r=R_{\rm ts})  & \text{if: } M_{\rm ej} < \frac{\dot{M} R_{\rm ts}}{v_{\rm w}}\, .
\end{cases}
\end{equation}

To obtain the time-averaged contribution of particle acceleration by remnants of SNe, one needs first to calculate the total amount of CR injected during the ED phase for a single SNR, that is  \cite{mitchell2024}:
\begin{equation}
\label{eq:CR_inj_ED}
    \mathcal{F}_{\rm snr}(t<t_{\rm st}) = \int_{0}^{t_{\rm st}} f_{\rm snr}(p) \frac{u_{\rm sh}}{4} \, 4\pi R_{\rm sh}^2 dt
    =  \frac{\pi}{3} \, R_{\rm st}^3 \, f_{\rm snr}(p)
\end{equation}
where $R_{\rm sh}(t) = R_{\rm st} (t/t_{\rm st})^{2/5}$, $u_{\rm sh}(t) = dR_{\rm sh}/dt = (2/5)\, R_{\rm sh}(t) /t$ and $t_{\rm st} = R_{\rm st}/u_{\rm ed}$. Then, the corresponding time-averaged contribution, spatially averaged over the bubble volume, is obtained by dividing the total injected particle spectrum by the bubble volume and multiplying by the number of supernovae that have exploded, yielding:
\begin{equation} \label{eq:f_snr_avg}
    \langle f_{\rm snr} \rangle =  N_{\rm sn}(t_{\rm esc}) 
    \frac{\mathcal{F}_{\rm snr}}{V_{\rm bubble}} 
    = \frac{N_{\rm sn}(t_{\rm esc})}{4} \, \frac{R_{\rm st}^3}{R_{\rm fs}^3} \, f_{\rm snr}(p)\,.
\end{equation}
We note that $N_{\rm sn}(t_{\rm esc})$ includes only those supernovae that have exploded within one advection time ($t_{\rm adv}=\int_{R_s}^{R_b} dr/u(r)$, where $u(r)= v_{\rm b} (R_{\rm ts}/r)^{-2}$ is the plasma velocity profile downstream the WTS). This ensures that only SNe whose accelerated particles are still confined within the bubble contribute to the time-averaged spectrum \cite{celli2024}.

Eq.~\eqref{eq:f_snr_avg} provides the time-averaged particle spectrum injected by SNRs, which means that it does not possess any radial dependence or account for propagation within the bubble. To address this, we treat particles accelerated by SNe as an additional contribution to those accelerated at the WTS. By doing so, the total CR content in the bubble ($f_{\rm cr}$) is obtained by replacing $f_{\rm ts}$ in Eq.~\eqref{eq:fcr} with $f_{\rm ts} + \langle f_{\rm snr} \rangle$:
\begin{equation}
\label{eq:fcr}
f_{\rm cr}(r, p) =
\begin{cases}
\left [f_{\rm ts}(p)+\langle f_{\rm snr}(p) \rangle \right ] \cdot \exp \left [ -\int_r^{R_{\rm ts}} \frac{v_{\rm w}}{D(r',p)}  dr' \right ]   & r < R_{\rm ts} \\[0.75em]
\left [f_{\rm ts}(p)+\langle f_{\rm snr}(p) \rangle \right ] \frac{ \left[e^{\alpha} + \beta(e^{\alpha_{\rm b}} - e^{\alpha}) \right]}{1+\beta(e^{\alpha_{\rm b}}-1)}  & R_{\rm ts} \leq r \leq R_{\rm b}~.
\end{cases}
\end{equation}

\section{Neutrino emission from the Galaxy}
\label{sec:nu}
The analysis we developed through this paper allows us to quantify for the first time the YMSC cumulative neutrino flux with a detailed model including both stellar winds and SNRs. We can therefore investigate what is their expected unresolved emission relative to the overall Galactic emission. 
In fact, the Milky Way Plane shines in high-energy non-thermal radiation produced by hadronic and leptonic interactions between CRs and target gas and/or radiation fields, which are responsible for the so-called diffuse emission. 
On top of such CR-induced emission, the same processes also take place inside and in the vicinity of acceleration sites, which have emerged as distinct $\gamma$-ray sources belonging to different populations, including YMSCs, SNRs, Pulsar Wind Nebulae (PWNe), microquasars, binary systems, etc. However, because high-energy non-thermal neutrinos are unambiguously associated with hadronic acceleration, they are key for the identification of CR sources. So far, all potential Galactic neutrino sources remain unresolved, including the most powerful Galactic $\gamma$-ray emitters observed by LHAASO \cite{lhaasoCat}. The role of unresolved Galactic neutrino sources was first investigated in \cite{ambrosone2024}, who concluded that current IceCube sensitivity is insufficient to distinguish them from the purely diffuse neutrino emission, thus stimulating us to realize novel template maps directly sensitive to YMSC flux.

In a recent work, \cite{mitchell2024} presented predictions for the neutrino flux from a sample of nearby YMSCs, selected from the \textit{Gaia} DR2 catalogue \citep{CantatGaudin20}, and limited to distances of $\lesssim$2~kpc. YMSCs  produce both $\gamma$-rays and high-energy neutrinos in the hadronic collisions of the accelerated protons with the target gas of the bubble, particularly from the shell region. The results of the investigation on the observed open cluster sample of the Milky Way indicate that only the most powerful clusters (e.g. Westerlund~1) may be detectable as neutrino resolved sources with a decadal long exposure. Given that 95\% of the Galactic cluster population is expected to lie at a distance $\gtrsim$2~kpc (see Fig.~\ref{fig:Dist_Size_Distrib}), we reasonably expect that most of the YMSCs will remain unresolved in terms of neutrino emission. As a consequence, the emission from these objects is likely to blend with the large-scale Galactic background.

We therefore proceed in Sec.~\ref{subsec:YMSCnu} to the computation of the neutrino emission from the YMSC population, while in Sec.~\ref{subsec:nuCR} we provide a description of the expected neutrino-induced emission by the CR interactions along the Plane. The sum of these two contributions represents our best prediction of the total neutrino expected flux from our Galaxy, which we are able to describe in terms of its energy and space distribution, under the assumption that particle accelerators other than YMSCs and their SNRs are negligible in terms of energetic proton production, as current $\gamma$-ray data tend to indicate. Later in Sec.~\ref{subsec:IC} we discuss comparison between current IceCube measurements of Galactic neutrino fluxes and our two-component (YMSCs plus CR-induced) diffuse model. Because the Galactic neutrino signal observed so far has emerged as a result of specific signal hypotheses, different than the one tested here, we warn the reader that this comparison is not expected to be quantitative by design. Nonetheless, it is useful to get a first insight into the level of contribution that the YMSCs of the Milky Way might overall provide. Our ultimate goal is to provide the community with novel templates of neutrino emission from the Galactic Plane, embedding up-to-date computation of neutrino sources, as we discuss in Sec.~\ref{sec:nuTemplate}. A joint fitting to available and future datasets of the source contribution, simultaneously with the CR-induced diffuse neutrino component, will be crucial to the emergence of Galactic neutrino sources, and therefore to the closure of the CR origin problem.

\subsection{Neutrino production by the YMSC population}
\label{subsec:YMSCnu}
The computation of the neutrino flux emerging from \emph{in-situ} hadronic collisions of the accelerated particles can be readily obtained once the CR and the target gas distributions are known. For a given YMSC, the all-flavour neutrino flux ($\phi_{\nu,\rm{sc}}$) is calculated as: 
\begin{equation}
\label{eq:PhiYMSC}
\phi_{\nu, \rm{sc}}(E_{\nu})=\frac{c}{4 \pi d_{\rm sc}^2} \sum_{i=e,\mu,\tau} \int_{E_\nu}^{\infty} \int_{0}^{R_{\rm fs}}  4 \pi r^2f_{cr}(r, E_p)n(r) \frac{d \sigma_{\nu_i}(E_p, E_{\nu})}{dE_{\nu}}\, dr\, dE_p \, ,
\end{equation}
where $d_{\rm sc}$ is the distance of the cluster from the Sun, $f_{cr}(r, E_p)=4\pi p^2\frac{dp}{dE_p}f_{cr}(r, p)$ is the CR distribution as a function of the proton kinetic energy ($E_p$), and the sum is performed over the three neutrino families. The functions $\frac{d \sigma_{\nu_i}}{dE_\nu}$ is the differential cross section for the production of neutrinos: we here adopt the AAFRAG parametrization provided in \cite{Kachelriess:2022khq}, which is based on QGSJET-II-04m interaction model.
The only other available parametrization for the neutrino cross-section is that of \cite{Kelner:2006tc}, based on SIBYLL 2.1, which yields a neutrino signal compatible with AAFRAG within $\sim 10\%$.

\begin{figure}
    \centering
    \includegraphics[width=\textwidth]{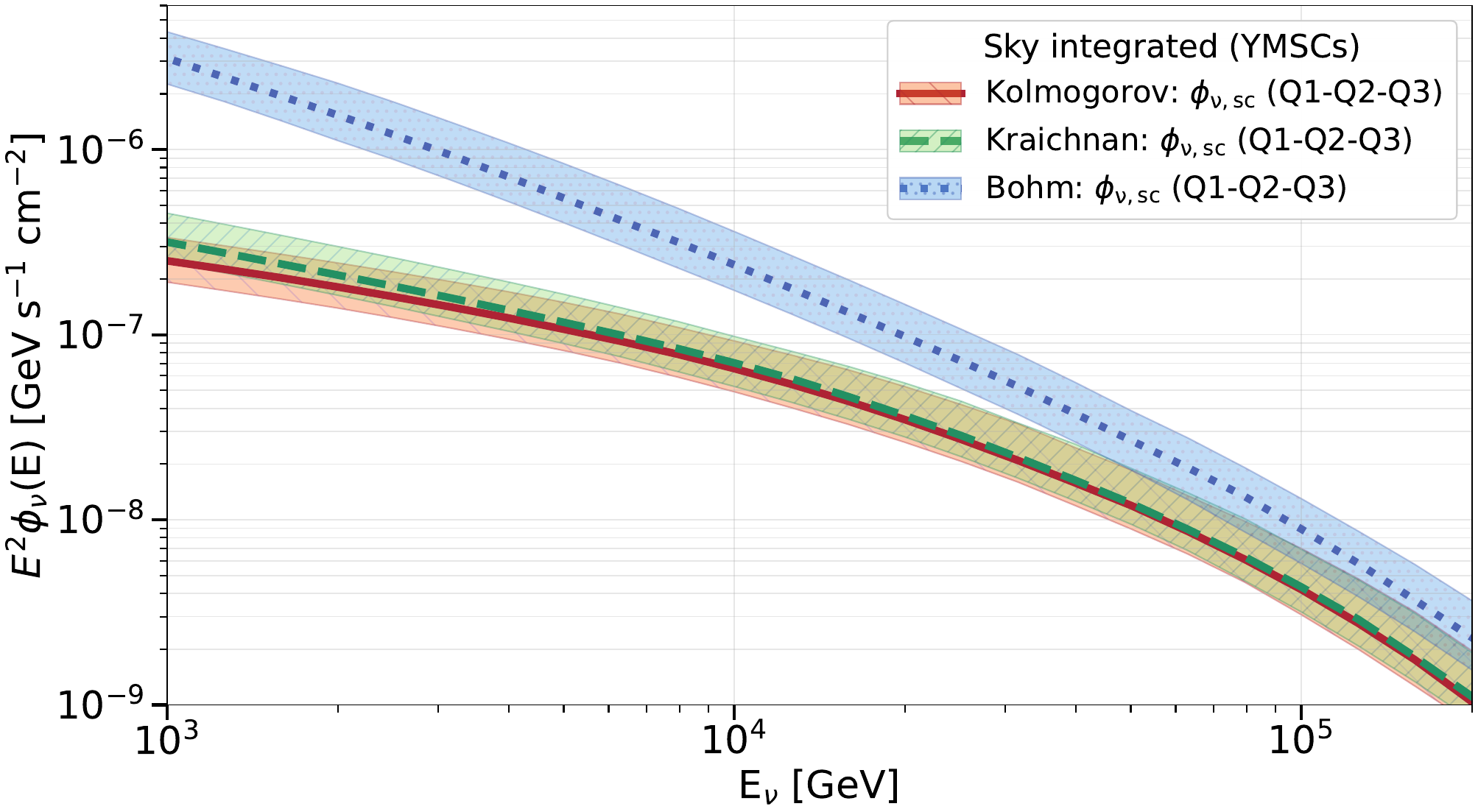}
    \caption{Predicted all-flavour, sky-integrated, neutrino flux from the population of Galactic star clusters, for the three different particle diffusion scenario considered. The spectra and their uncertainties are derived from 100 different realizations of the Galactic cluster population. The lines represent the median flux (Q2), while the hatched bands are the 25th (Q1) and 75th (Q3) percentiles.}
    \label{fig:nuFluxes_SC}
\end{figure}

The resulting cumulative all-sky and all-flavour neutrino emission from the Galactic population of YMSCs, as obtained by summing the contribution from the 2243 objects expected in the Milky Way, is shown in Figure~\ref{fig:nuFluxes_SC} for the three different particle diffusion scenarios here considered. The total neutrino spectra and their associated uncertainties are derived from 100 different realizations of the Galactic cluster population, using the median flux together with the 25th and 75th percentiles to define the uncertainty band. The most effective scenario in particle diffusion, namely the Bohm domain, produces the largest neutrino fluxes. 
 
\subsection{Galactic CR-induced diffuse neutrinos}
\label{subsec:nuCR}
We here present the method and assumptions used to calculate the CR-induced diffuse neutrino emission from hadronic collisions with target gas in the Galactic Plane. To this extent, we adopt the model described in \cite{Vecchiotti:2024kkz} for $\gamma$-rays and extend it to the neutrino case. Analogously to Eq.~\eqref{eq:PhiYMSC}, the all-flavour diffuse neutrino emission as a function of the neutrino energy $E_{\nu}$ and arrival direction $\hat{n}_{\nu}$ can be calculated as \cite{Pagliaroli:2016, Cataldo:2019qnz}:
\begin{eqnarray} 
\label{eq:PhiNu_sea}
\varphi_{\nu,{\rm diff}}(E_{\nu},\hat{n}_{\nu}) = 
\sum_{i=e,\mu,\tau} \int_{E_{\nu}}^{\infty} dE_p\; \frac{d \sigma_{\nu_i}(E_p,E_{\nu})}{dE_{\nu}} 
\int_{0}^{\infty} d\ell\; \varphi_{CR}(E_p ,r_{\odot} + \ell \hat{n}_{\nu})\, n_{\rm H}(r_{\odot} + \ell\hat{n}_{\nu})   
\end{eqnarray}
where $\varphi_{\rm CR}(E_p,{\bf r})$ is the CR flux as a function of the CR particle energy and position in the Galaxy ${\bf r}$. Here, ${\bf r} =  {\bf r_{\odot}} + l \hat{n}_{\gamma}$, with ${\bf r_{\odot}}=8.5$~kpc being the position of the Sun with respect to the Galactic Center, and $\ell$ being the distance along the line of sight.
The integral is performed over the nucleon energy $E_p$ and along the line of sight $\ell$.
The modelling of the diffuse emission is subject to significant uncertainties \citep{Schwefer:2022zly}, associated with the quantities listed above, therefore, we consider different modelling hypothesis. 
For the gas distribution, we consider the model provided by the \texttt{GALPROP} code \cite{Galprop}, which includes contributions from both atomic hydrogen, ${\rm H}$\textsc{i}, and molecular hydrogen, $\rm{H}_{2}$, traced by the ${\rm H}$\textsc{i} \cite{Kalberla2005,hi4pi2016} and the CO emission line \cite{Dame2001}, respectively. A uniform $X_{\rm{CO}}$ conversion factor of $1.9\times 10^{20}\, \rm cm^{-2}\, K^{-1}\, km^{-1}\, s$ is assumed to relate CO brightness to the H$_2$ column density, while the ${\rm H}$\textsc{i} is calibrated assuming a uniform spin temperature of $125$ K \cite{Strong2004}. We consider this gas distribution as our reference model for the presentation and discussion of the results, while in Appendix \ref{app:C} we discuss the impact of adopting a different gas distribution: in particular, we further test a more recent model based on \cite{Soding_2025}, providing three-dimensional maps of both atomic and molecular hydrogen and incorporating updated observational constraints and a refined reconstruction of the gas distribution. Both gas model are multiplied by a factor 1.42 to account for the presence of heavier elements in the target gas. This factor reflects the Solar System composition, assumed to be representative of the entire Galactic Disk \citep{Ferriere:2001rg}.

In turn, the CR flux is parametrized as:
\begin{equation}
\varphi_{\rm CR}(E_p,{\bf r}) = \varphi_{\rm CR,\odot}(E)\,g({\bf r})\,h({E_p,\bf r})\ ,
\label{Eq:CR_flux}
\end{equation}
where $\varphi_{\rm CR,\odot}(E_p)$ is the nucleon flux measured at the Sun position, while $g({\bf r})$ describes the spatial distribution of CRs throughout the Galaxy.  
The function $g({\bf r})$ is dimensionless and normalized to 1 at the Sun position, ${\bf r}_\odot$.
For it, we adopt the same parametrization as in \cite{Cataldo:2019qnz}, where it is defined as the solution of a 3D isotropic diffusion equation with constant diffusion length, $R$, set to infinity\footnote{This value best reproduces the $\gamma$-ray emissivity data at $20$ GeV provided by {\it Fermi}-LAT \cite{Cataldo:2019qnz}.}, and assuming stationary CR injection proportional to the SNR number density, as parametrized by \cite{Green:2015isa}.
This spatial distribution peaks at $\sim 5$ kpc from the Galactic centre and reflects the distribution of CR sources in our Galaxy.
The function $h({E_p,\bf r})$ in Eq.~\eqref{Eq:CR_flux} accounts for the possibility that the CR spectral index varies with Galactocentric distance. In particular, it incorporates the spectral hardening of large-scale $\gamma$-ray emission in the inner Galaxy, as inferred from analysis of the {\it Fermi}-LAT data \cite{Acero:2016qlg, Yang:2016jda, Pothast:2018bvh}. Specifically, $h({E_p,\bf r})$ is defined as:
\begin{equation}
h(E_p,{\bf r})=\left(\frac{E_p}{\overline{E}}\right)^{\Delta({\bf r})}
\label{Eq:h_funct}
\end{equation}
where the pivot energy $\overline{E}$ is set at $20$ GeV,  corresponding to the energy at which spectral hardening was inferred from {\it Fermi}-LAT data \cite{Acero:2016qlg, Yang:2016jda, Pothast:2018bvh}, while the index variation $\Delta({\bf r})$ in Galactic cylindrical coordinates reads as:
\begin{equation}
  \Delta(r,z) =\Delta_0
  \begin{cases}
    \left(1 - \frac{r}{r_{\odot}} \right) & r\le 10\, \rm kpc \\
    \left(1 - \frac{10\, \rm kpc}{r_{\odot}} \right) & r > 10\, \rm kpc
  \end{cases}
\label{Eq:Delta} 
\end{equation}
with $\Delta_0=0.3$ being the difference between the CR slope at the Galactic Center and its value at the Sun position. 
This approach is essentially equivalent to the ``space-dependent" model from \cite{Lipari:2018gzn}, and consistent with the results obtained from the KRA$_\gamma$ CR propagation by \cite{Gaggero:2014xla, Gaggero:2015, Gaggero:2017jts}, the only difference being that we allow for a further softening of the CR index beyond $r=8.5$ kpc.
As a result, our longitude‑averaged diffuse emission spectrum is softer than $2.5$, the averaged index of the KRA$_\gamma$ model displayed in comparison with the IceCube Galactic diffuse emission data \cite{IceCubeScience}.

\begin{figure}
    \centering
    \includegraphics[width=\textwidth]{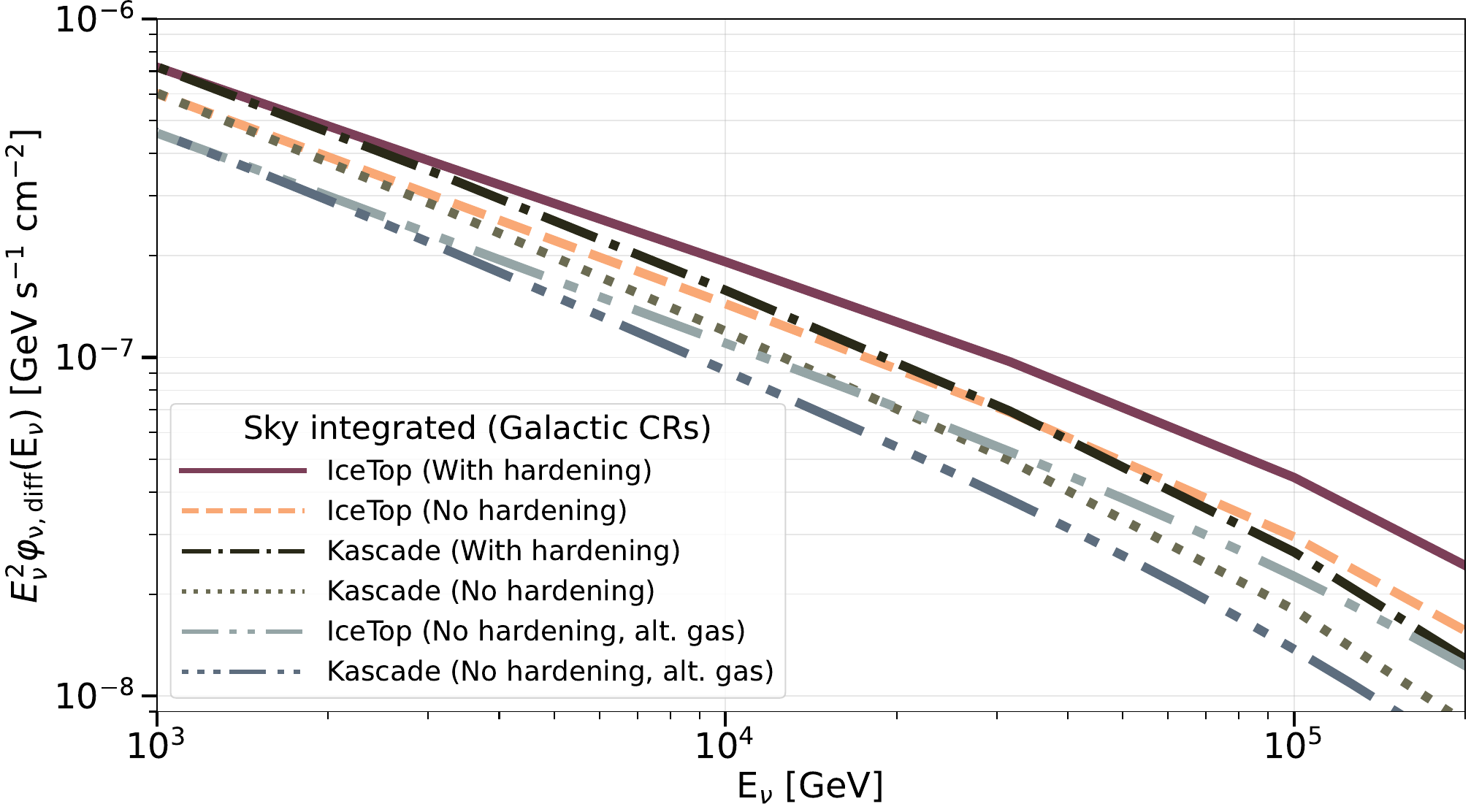}
    \caption{Predicted all-flavour, sky-integrated, diffuse neutrino flux from the Galactic CR sea. A solid purple line and a dashed orange line show the emission based on the IceTop parametrization \citep{IceCube:2019hmk}, with and without spectral hardening. A dot-dashed black line and a dotted grey line show the emission based on the KASCADE parametrization \citep{Vecchiotti:2024kkz}, with and without spectral hardening. Finally, a double-dot-dashed light grey line and a triple-dot-dashed blue line show the emission based on the IceTop and Kascade parametrization respectively but using the alternative gas model provided by \cite{Soding_2025}.}
    \label{fig:nuFluxes_CRsea}
\end{figure}

Uncertainties in the local nucleon CR spectrum $\varphi_{\rm CR,\odot}(E_p)$, specifically chemical composition and energy position of the \emph{knee}, reflect into uncertainties upon the expected secondary radiation emerging from their collisions.
We here consider two possible data-driven parametrization: the first one, obtained from \cite{Dembinski:2017}, includes the contribution from all the nuclear species and the proton spectrum in chosen to reproduce the data from IceTop \cite{IceCube:2019hmk}. The second parametrization is instead obtained by substituting the aforementioned proton spectrum with an alternative one provided in \cite{Vecchiotti:2024kkz},  which reproduces the KASCADE measurements \cite{Apel:2013uni}.  Fig.~\ref{fig:nuFluxes_CRsea} shows the expected all-flavour and sky-integrated neutrino-induced emission from the Galactic CR sea for the two chosen CR parametrizations, namely with and without the hardening effect towards the Galactic Center, as well as for both the reference (\texttt{GALPROP}) gas template and the alternative one based on \cite{Soding_2025} (the latter is shown for simplicity only in the no-hardening scenario).

\subsection{Comparison with neutrino observations from the Galactic Plane}
\label{subsec:IC}
Neutrino emission from the Galactic Plane has been reported by IceCube in a highly sensitive search exploiting 10 years of data, capable of identifying a flux of events a factor 10 about less intense than the overall cosmic diffuse neutrino emission \cite{IceCubeScience}. The Galactic signal was observed in the event sample characterised by a cascade-like topology, whose selection with respect to track-like events benefits of a reduced atmospheric neutrino background at the expenses of angular resolution, thus favouring the observation of diffuse emissions.
Three model dependent analyses have been employed, based upon predictions from the so-called \emph{$\pi^0$} \cite{fermi2012} and KRA$_\gamma$ \cite{kraGamma} models. Both models have been tuned using GeV $\gamma$-ray data, but they adopt different CR distributions, hence the resulting predictions for the spectral and spatial distribution of the expected neutrino events produced by hadronic collisions are different. Specifically, the former assumes a uniform CR distribution with gas and dust maps from \texttt{GALPROP} \cite{galprop2002}, resulting into a differential energy spectrum of Galactic diffuse photons from $\pi^0$-decay distributed as $dN/dE \propto E^{-2.7}$ at GeV energies, which is extrapolated with this very same unbroken power-law spectrum up to the TeV energy range for the purposes of neutrino analysis. The KRA$_\gamma$ model, in turn, is a more refined calculation of the diffuse Galactic emission based upon the CR transport code \texttt{DRAGON} \cite{dragon}, featuring a radial dependence of the CR spectrum, putatively ascribed to a spectral hardening of the diffusion coefficient towards the inner Galaxy, as to reproduce $\gamma$-ray observations above 10~GeV reported by Fermi-LAT \cite{fermi2016}. As such, the KRA$_\gamma$ model can exploit specific features of the CR spectrum, e.g. with regard to mass composition and energy breaks. In the IceCube analyses of the Galactic Plane \cite{IceCubeScience}, two different assumptions have been considered with regard to the location the CR-proton knee, at either 5~PeV or 50~PeV energies, respectively in the so-called KRA$^5_\gamma$ and KRA$^{50}_\gamma$ models\footnote{Several updates of the KRA$_\gamma$ model are currently available, with minimal and maximal models bracketing available CR data from different observatories across the knee \cite{pedro2023}. They also exploit a different gas distribution than previous releases.}. The gas distribution adopted in these KRA$_\gamma$ templates is the same as in \texttt{GALPROP} \citep{galprop2002}. 
Fitting these templates to the neutrino data provided the signal flux normalization that is most consistent with observations. The largest significance above background fluctuations is currently provided at $4.5\sigma$ level, with a slight preference towards the \emph{$\pi^0$} template. Preliminary investigation with the addition of the track-like sample has been reported to increase the signal significance of the IceCube data selection beyond the 5$\sigma$ threshold \citep{ICtracks2025}.  
The cascade-only best fit results indicate a different normalization than what the various templates predicted: specifically IceCube data point towards a higher Galactic neutrino flux than the \emph{$\pi^0$}-model (about a factor 5 higher), while a lower Galactic neutrino flux than expected is preferred in the KRA$^5_\gamma$ template fitting (a factor 2 smaller). A possible explanation of the disagreement might reside in an additional contribution from Galactic neutrino sources or in a different location of the CR knee than what has been assumed. 

The possible presence of a cumulative contribution from Galactic sources was additionally tested in IceCube data, exploiting different source populations, including SNRs, PWNe, and unidentified TeV emitters \cite{IceCubeScience} via catalog-based stacking searches. 
However, the current statistics results insufficient to discriminate between the purely diffuse and the diffuse plus source hypothesis: thus, so far, Galactic neutrino sources remain to be unveiled. The negative result from the point-source analysis is not surprising, when considering the severe limitations of the methods adopted, e.g. the extended spatial overlap among the catalog objects and regions predicting the largest neutrino fluxes in the CR-induced component, as well as the uniform spectral hypothesis concerning the sources involved. 

\begin{figure}
    \centering
    \includegraphics[width=\textwidth]{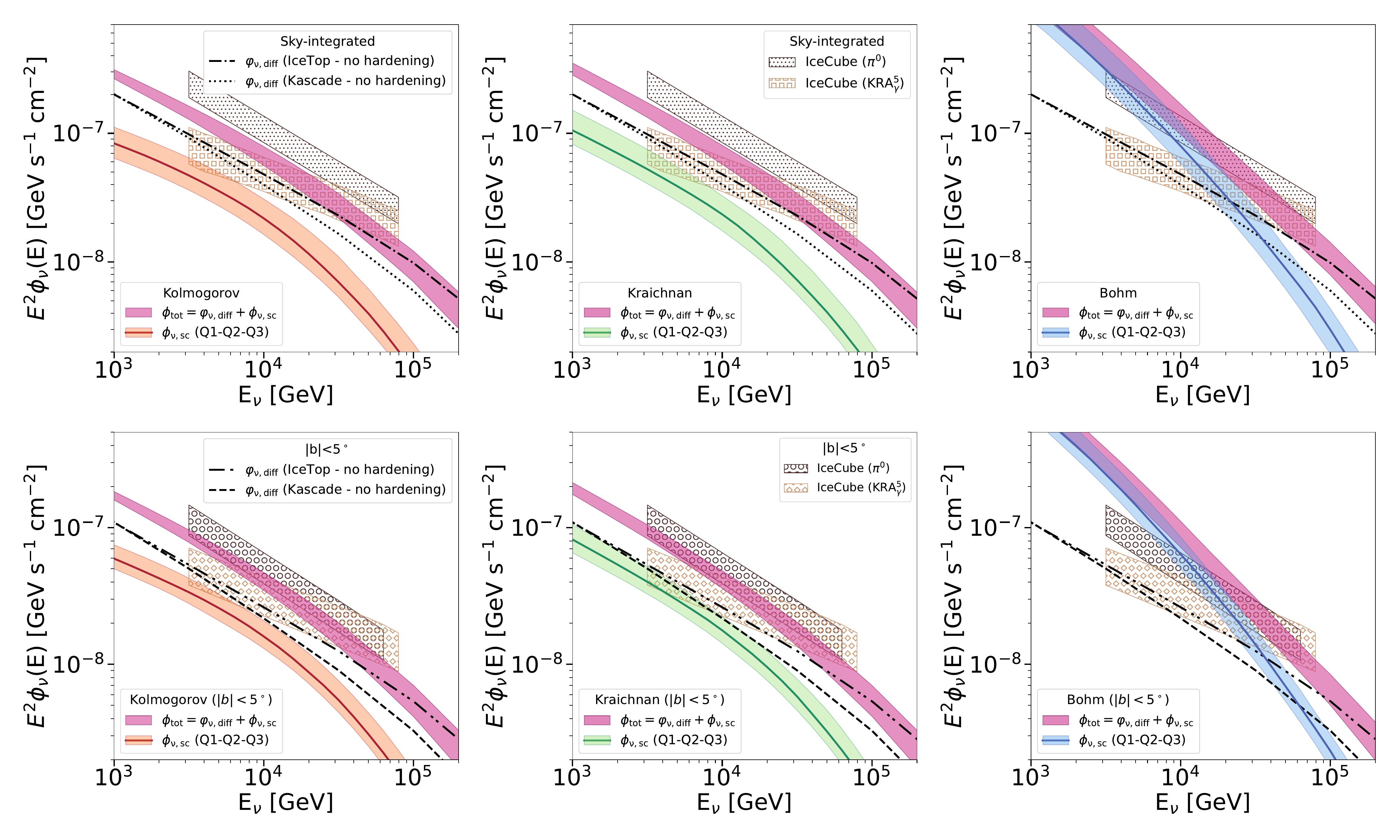}
    \caption{Predicted all-flavour  neutrino energy fluxes from the Galactic Plane: the individual contribution by YMSCs and by the CR sea is shown separately, as described in the legend. Diffusion domain is assumed to be Kolmogorov in the left panels, Kraichnan in the middle panels, and Bohm in the right panels. Top panels refer to the all-sky expected emission, while bottom panels limit the spatial integration to $|b|<5^\circ$. The observed Galactic neutrino emission by IceCube is also shown for comparison, respectively with black and brown dotted bands in the $\pi^0$ and KRA$_\gamma^5$ fitting models, with bands reporting the $1\sigma$ uncertainty level.}
    \label{fig:nuFluxes}
\end{figure}

\begin{figure}
    \centering
    \includegraphics[width=\textwidth]{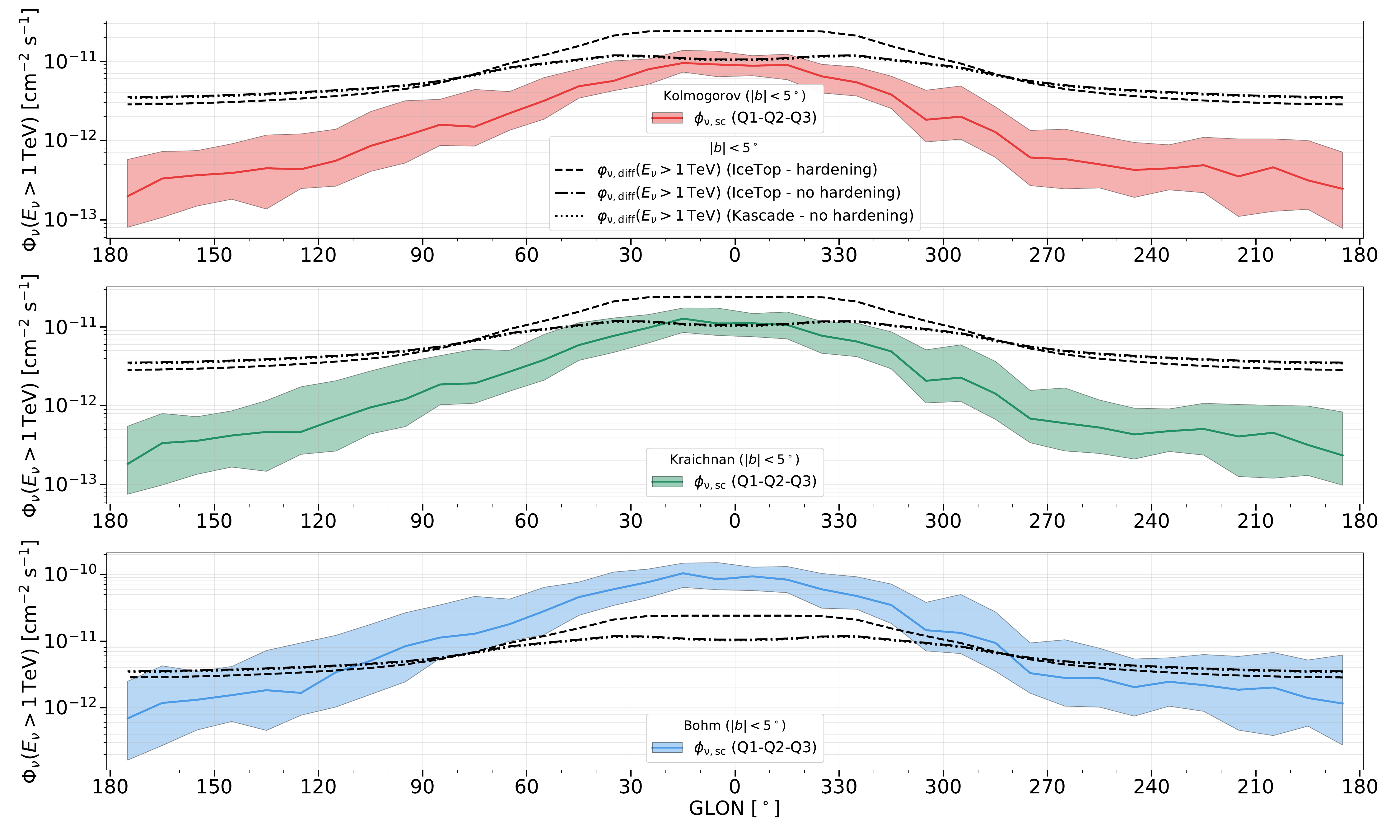}
    \caption{Longitudinal profiles of the expected all-flavour neutrino fluxes above 1~TeV, integrated in the latitude range $|b|<5^\circ$. The contribution by YMSCs is provided by the coloured lines, with the diffusion domain assumed in the modelling being Kolmogorov in the top panel, Kraichnan in the middle panel, and Bohm in the bottom panel. Dot-dashed and dotted lines in each panel correspond to the CR-induced neutrino emission, reproducing respectively CR data by IceTop and Kascade with no hardening in the inner Galactic region. The dashed line corresponds instead to the CR-induced neutrino emission, reproducing IceTop CR data under the assumption of a spectral hardening in the inner Galactic region. Differences between the IceTop and KASCADE cases (in the no hardening scenario) are negligible, since the emission is dominated by the lower threshold of integration (here set at 1~TeV), whereas the models mainly diverge around PeV energies.}
    \label{fig:longProfNu}
\end{figure}

In these regards, our approach here is more effective than catalog-based searches as it deals with the diffuse unresolved emission by the overall population of neutrino sources, schematically described as dominated by YMSCs and their SNe, rather than dealing with specific emitters. Therefore, in Fig.~\ref{fig:nuFluxes} we show the total neutrino fluxes expected in our model, obtained as the sum between the YMSC component (together with the SNe exploding therein) and the CR-induced one in the scenario without the radial-dependent hardening, and compare them to IceCube detected neutrino fluxes from the Galactic Plane, both in the all-sky analysis and limited to the latitude range $|b|\leq 5^\circ$ (where sources are expected to be of larger relevance). The YMSC contribution is shown separately from the purely diffuse neutrino emission due to the CR sea, in order to assess its relevance under different assumptions about particle diffusion within the sources. 
In all cases except Bohm, we observe that the total contribution from YMSCs and CR-induced neutrino flux has a spectral shape more consistent with the $\pi^0$ than the KRA$_\gamma$ model, while for Bohm diffusion there's not much spectral resemblance with any of the tested templates.
We further recall that the neutrino emission from YMSCs is calculated assuming the standard 10\% CR acceleration efficiency from both stellar winds and SNe\footnote{Because the normalization of the neutrino flux from YMSCs scales linearly with the assumed CR acceleration efficiency, fitting the spatial and spectral templates from this work will provide constraints to this parameter.}. With such an assumption, the Bohm diffusion case results to be the only one where the YMSCs dominate over the CR-induced component, at least below $E_\nu \sim 50$~TeV. In less efficient diffusion domains, neutrinos from YMSCs constitute a limited fraction of the measured diffuse emission from the Galactic Plane: specifically, in the Kraichnan case, the neutrino flux from $|b|\leq 5^\circ$ receives comparable contributions by YMSCs and CR interactions, with the latter component dominating at the highest energies. 
This case also appears to better reproduce the IceCube fitted templates, similar to the predictions in the Kolmogorov scenario from the same sky region, i.e., the left and middle bottom panels of Fig.~\ref{fig:nuFluxes}.
The longitudinal profiles of the expected neutrino emission from $|b|\leq 5^\circ$ are reported in  Fig.~\ref{fig:longProfNu}, again separately for the CR-induced component and the YMSC one.
From these panels, it clearly emerges that the inner longitude range from the Galactic Center, $|l|\leq 30^\circ$ (also known as Galactic Ridge), contains most of the YMSC expected emission, while larger longitudes are always dominated by the CR-induced emission, regardless of the CR spectrum being more consistent with IceTop or KASCADE results. 
In this region, the median ratio between the YMSC and CR-induced contributions without hardening is $\sim 0.74$, $\sim 0.94$, and $\sim 7.48$ for the Kolmogorov, Kraichnan, and Bohm cases, respectively.

\begin{figure}
    \centering
    \includegraphics[width=\textwidth]{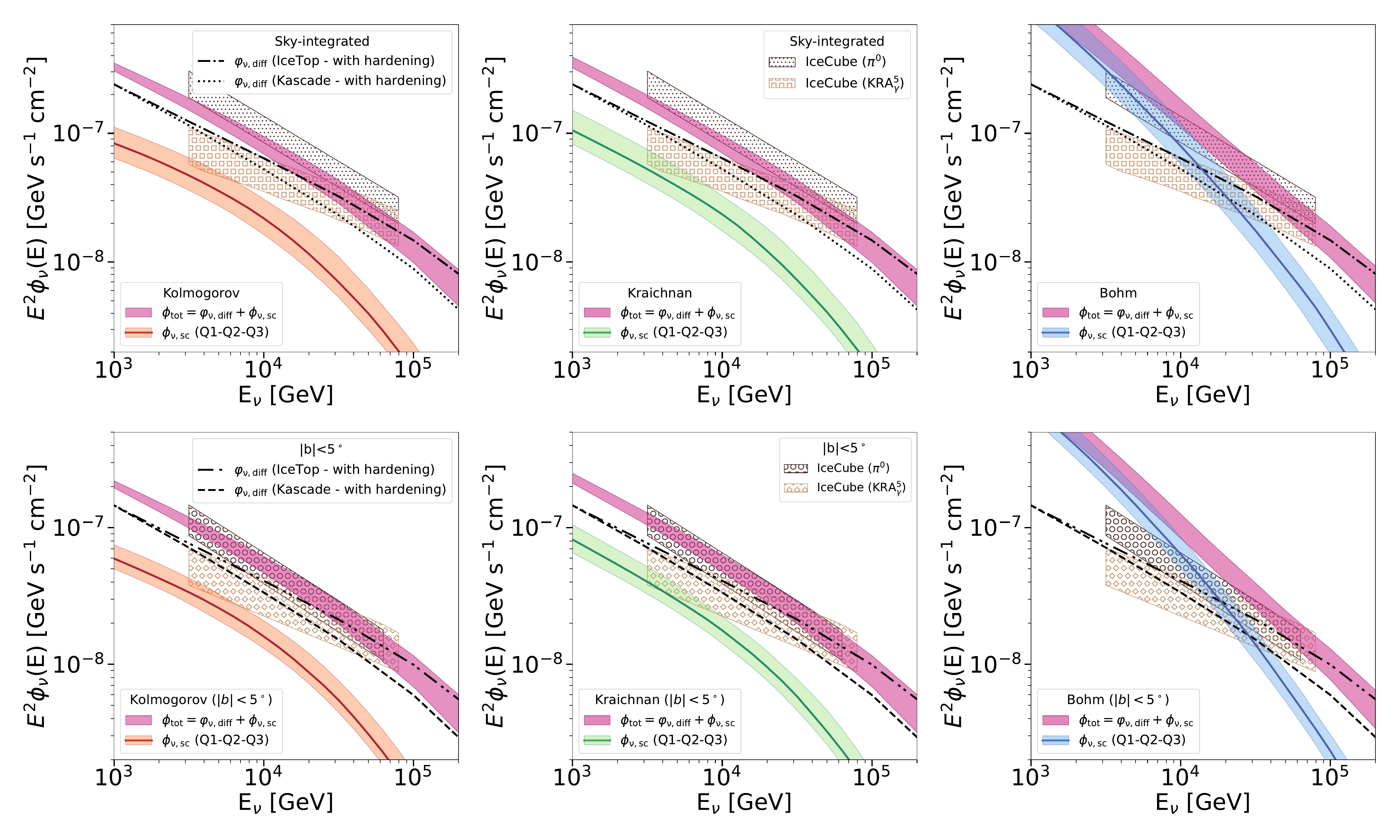}
    \caption{Same as Fig.~\ref{fig:nuFluxes}, this time considering the CR diffuse emission with a hardening in the spectral index in the inner Galactic region.}
    \label{fig:nuFluxes_Hard}
\end{figure}

Finally, Fig.~\ref{fig:nuFluxes_Hard} shows the results obtained when including a spectral hardening of the CR sea in the inner Galactic  region. Compared to the case without any hardening, the relative contribution of YMSCs in the Kolmogorov and Kraichnan scenarios is reduced, lying a factor of $\sim 2.5$ below the diffuse CR-induced component at $E_\nu \sim 10$~TeV. The same behaviour is reproduced in the corresponding longitudinal profiles, shown in Fig.~\ref{fig:longProfNu}, where the diffuse CR-induced emission becomes the dominant contribution in the Galactic Ridge. In this region, the median ratio between the YMSC and CR-induced neutrino flux decreases to $\sim 0.37$ and $\sim 0.47$ for the Kolmogorov and Kraichnan diffusion scenarios, respectively. The Bohm case, on the other hand, remains source-dominated, with a median YMSC-to-diffuse ratio of $\sim 3.75$.
Nevertheless, in the Kraichnan and Kolmogorov scenarios, the total neutrino spectrum remains in good agreement with IceCube data, while the Bohm case tends to overpredict the flux below $E_\nu \sim 10$~TeV. Overall, the inclusion of a hardening in the CR spectrum leads to a better agreement with the observed flux at $E_\nu \sim 100$~TeV compared to the scenario without hardening.

Finally, to validate our methodology, we compute the diffuse $\gamma$-ray emission produced by YMSCs, and compare the resulting predictions with current measurements of the Galactic diffuse $\gamma$-ray emission to ensure full consistency, similarly to \cite{Menchiari25}. We find that our model remains compatible with the existing $\gamma$-ray observations. This comparison, presented in Appendix~\ref{sec:appA}, provides a cross-check for our neutrino predictions, while also offering an independent estimate of the contribution of unresolved YMSCs to the non-thermal emission of the Milky Way.

\subsection{Novel templates to search for Galactic neutrino sources}
\label{sec:nuTemplate}
\begin{figure}
    \centering \includegraphics[width=\textwidth]{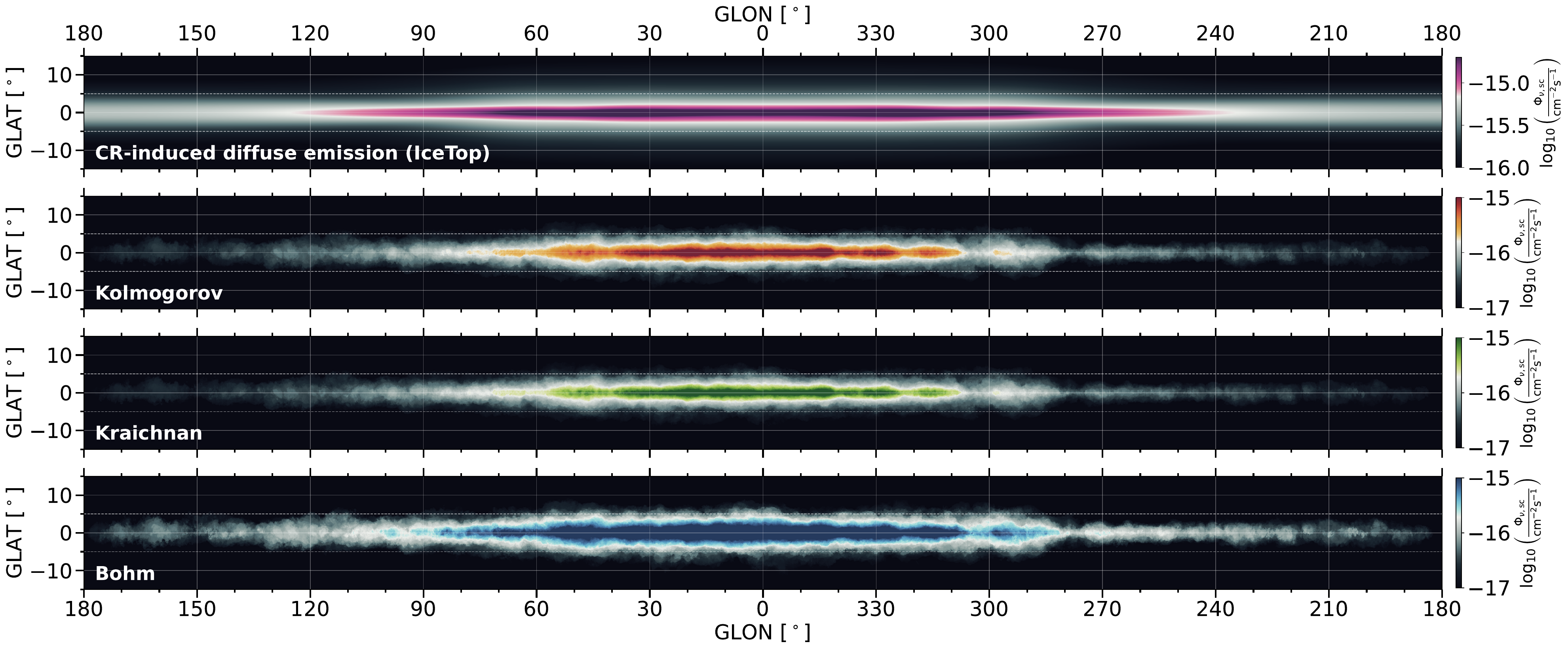}
    \caption{Predicted sky maps of the all-flavour neutrino flux above 1 TeV. The top panel shows the CR-induced neutrino emission, in accordance with IceTop CR data at the knee and without spectral hardening in the inner Galactic region, followed by the YMSC neutrino emission in different assumptions of diffusion domain, according to the legend.}
    \label{fig:nuTemplates}
\end{figure}

The previous discussion was restricted to a qualitative level, because the IceCube observations are tuned on specific template models which differ from ours in many regards, as well as among themselves, such that below $E_\nu \sim 100$~TeV the disagreement among the fitted emission with the $\pi^0$ and KRA$_\gamma$ models leaves the Galactic neutrino flux basically unconstrained. 
Not only are the spectral assumptions adopted in IceCube analyses different from ours, but they are also intended to describe only the CR-induced neutrino emission from the Galactic Plane, with no contributions from hadronic sources. 
Because our predictions go beyond existing model, we encourage IceCube and other  collaborations operating large volume neutrino telescopes to consider implementing our templates in future template fitting analyses, as to independently constrain the purely diffuse flux from the source component with a physically motivated and up-to-date model.
To this extent, we provide separately the source and diffuse emission templates in the online material of the paper. 
As such, we show in Fig.~\ref{fig:nuTemplates} the expected sky distribution of the neutrino events, separated in its different constituents. The source templates are constructed by considering, for each line of sight, the median neutrino flux from YMSCs, calculated across 100 different realizations of the Milky Way. The emission from each YMSCs is modelled as a uniform disk, with a size equal to the projected bubble radius.
With regards to the CR-induced component, the template is constructed straightforwardly by calculating Eq.~\ref{eq:PhiNu_sea}. In Fig.~\ref{fig:nuTemplates}, we show only the template referring to IceTop data with no hardening,  as a reference.
We notice that, because the emission shown is integrated above 1 TeV, while the differences among the IceTop and Kascade CR models are located around the CR knee region, no major visual variation is found among them. 
For the cases with spectral hardening, the specific templates are not shown here but they are made available in the online material of the paper.

Concerning the YMSC component, we notice that it appears to be concentrated in $|b|\leq 5^\circ$. Therefore, this latitude range should be regarded as the preferable one for analyses aiming at the emergence of neutrino sources, also in terms of the spectral comparison shown in Fig.~\ref{fig:nuFluxes}. 
In these regard, the adoption of track-like events with angular resolution well below that adopted by IceCube in the cascade-based analysis will be decisive for Galactic neutrino source detection, thanks to the reduced background in the coincident spatial region. 
To this purpose, we explicitly report in Appendix \ref{app:B} the smeared template sky maps separately for the track and cascade channels, as obtained according to the angular resolution of the currently major Cherenkov neutrino telescopes of the Northern and Southern hemisphere, KM3NeT and IceCube respectively (see Fig~\ref{fig:nuTemplates2}).

\section{Conclusions}
\label{sec:conclusions}
The recent detection of high-energy neutrinos from the Galactic Plane has been a remarkable achievement, while not yet sufficient to clearly identify hadronic accelerators in the Milky Way. 
The limited angular resolution of the cascade event sample adopted in IceCube analyses favoured the prevalence of the guaranteed CR-induced neutrino diffuse component over the (also expected) neutrino source term.  
To overcome this limitation, we have developed a novel description of the expected neutrino emission from the Galactic Plane, additionally including the cumulative contribution of major Galactic CR sources, namely YMSCs powered by both stellar winds and SNRs. 
The model is built upon up-to-date description of CR acceleration and transport in these systems, as well as realistic modelling of the YMSC spatial and age distribution in our Galaxy. 
The comparison between the spectral energy distributions of our models with IceCube observations of the Galactic Plane suggests that the Bohm diffusion scenario for particles inside YMSCs would overproduce neutrinos, while Kraichnan and Kolmogorov cases appear to provide more suitable propagation regimes for the accelerated particles. 
As such, it is crucial to directly test our models against available and future neutrino datasets to derive a first measurement (or constraints) of the CR acceleration efficiency of these objects. The quality of observations, and therefore their agreement to models, are expected to improve in analyses with track-like events, thanks to their superior angular resolution: in these regards, water-based Cherenkov neutrino telescopes in the Northern Hemisphere as KM3NeT, having  a privileged view of the Galactic Plane almost free from the atmospheric muon background, are expected to finally open the way to the investigation of hadronic CR sources in our Galaxy.

\appendix

\section{Galactic neutrino template maps: expectations by KM3NeT and IceCube}
\label{app:B}
\begin{figure}
    \centering 
    \includegraphics[width=\textwidth]{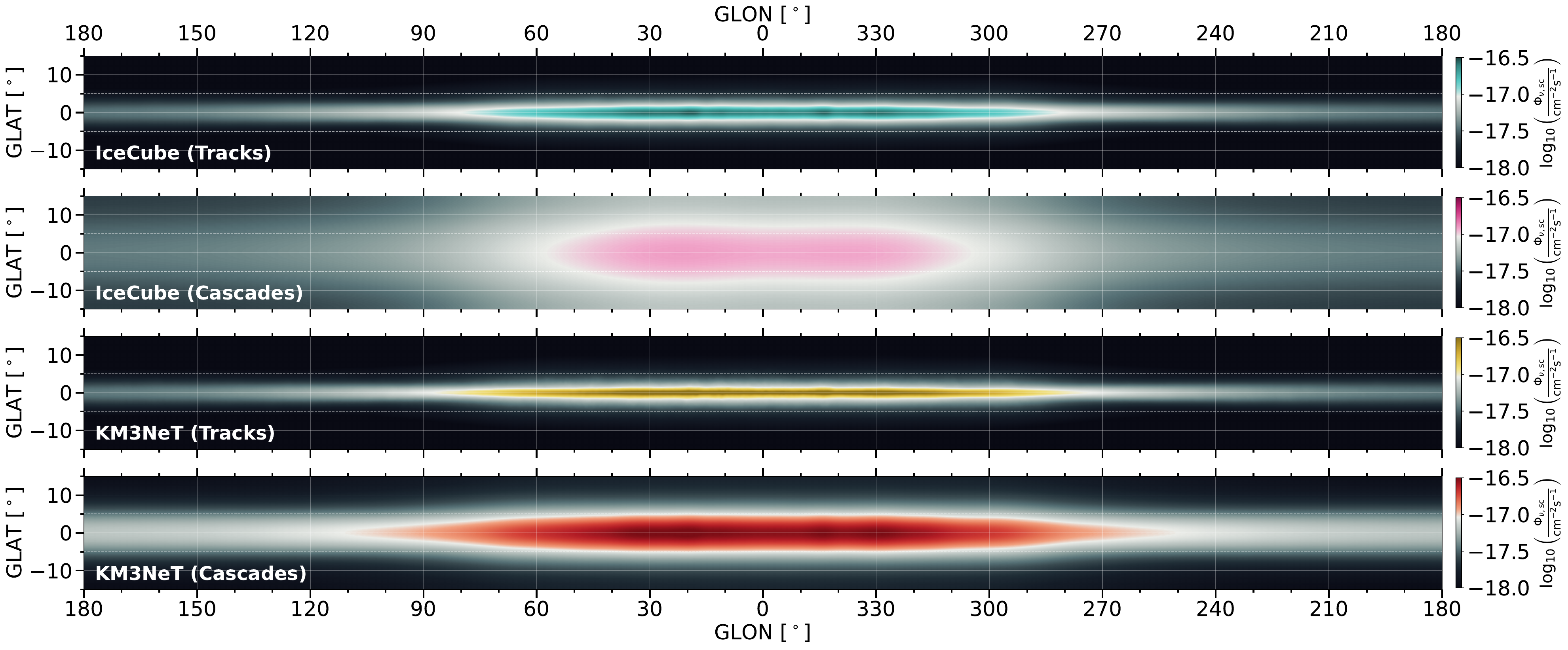}
    \caption{Total (CR-induced emission plus YMSC contribution, assuming the Kraichnan scenario) integrated Galactic neutrino flux maps for $E_\nu > 10 \text{ TeV}$. The two upper panels show the expected signal in the IceCube experiment, respectively in the track and cascade channels, while the two lower panels provide the same signal in KM3NeT, as indicated in the labels.}
    \label{fig:nuTemplates2}
\end{figure}

In this Appendix, we show our predicted Galactic neutrino template maps as they are expected to be observed by the neutrino telescopes IceCube and KM3NeT \cite{km3loi}, separately for the track and cascade channels, to clearly show the distinction of these two event samples and the expected improvements with future analyses (see Fig.~\ref{fig:nuTemplates2}). \\ 
IceCube, being located at the South Pole, observes the Southern sky via downward-going events, and therefore it relies on starting events to suppress the huge atmospheric background, at the expense of the effective volume adopted in these analyses compared to the instrumented one. 
We recall that the current detection of Galactic neutrinos has occurred in IceCube via deep neural network (DNN) reconstruction of the cascade channel \cite{IceCubeScience}. This event sample was specifically adopted in the Galactic Plane search to take advantage of its reduced background compared to track-like events; despite of a worse angular resolution, the energy deposition of cascade-like events results in an almost calorimetric measurement of the primary neutrino energy, being more concentrated than that of tracks, which makes them the ideal event sample for the detection of diffuse fluxes. 
Currently, more analyses are ongoing to exploit the contribution of additional event samples, including the so-called Enhanced Starting Track Sample (ESTES), in order to achieve the maximal sensitivity to the Galactic flux \cite{ICestes}. 
Each event sample features different selection efficiencies, background contamination, and sky exposure. Therefore, investigations are currently underway to determine the most sensitive event sample for Galactic Plane neutrinos in IceCube \cite{iceman}. \\ 
On the other hand, Northern hemisphere instruments  can observe the Galactic Plane via the high-purity sample of upward-going events, profiting from the filtering effect provided by the Earth to remove the atmospheric muon background. 
This allows the inclusion of through-going track-like events in the Galactic Plane analyses, with a large exposure and an optimal angular resolution. 
These properties will enable to get both a much more resolved view of the neutrino emission from our Galaxy and, most importantly, will favour the emergence of Galactic neutrino sources thanks to the reduced background rate in the angular search region. 
The ANTARES neutrino telescope in the Northern hemisphere, which collected data for 15 years, investigated the Galactic Ridge and found a mild ($\sim 2\sigma$) excess of events compared to the expected background \cite{ANTARESgr}, equivalent to a neutrino flux from this area of $\sim 5\times 10^{-8}$~GeV$^{-1}$~cm$^{-2}$~s$^{-1}$ at 40~TeV  (consistent with IceCube measurements from the extended sky area of the entire Galactic Plane).
The significance of the ANTARES signal did not improve with analyses using template-fitting procedures similar to that employed by IceCube \citep{ANTARES:2025wvi}.
The high-energy KM3NeT detector, so-called ARCA, in its final two building-block configuration is foreseen to instrument a volume 100 times larger than ANTARES, thus providing significantly improved results about the neutrino signal from the Galactic Plane. \\
To illustrate the key role of instrument location and event sample adopted in the analysis, we proceed by first splitting our all-flavour Galactic template map, as shown in Fig.~\ref{fig:nuTemplates}, into its expected track-like and shower-like components. 
They amount to $\sim 30\%$ and $\sim 70\%$, respectively, as it results from assuming equal neutrino flavour composition at Earth in long baseline oscillations from pure pion production\footnote{We accounted for the fact that $\sim 22$\% of $\nu_\mu$ interactions proceed via the exchange of the neutral $Z^0$ boson, resulting in cascade-like events. Moreover, since $\sim 70$\% of $\nu_\tau$ interactions proceed via charged-current interactions and the branching ratio of $\tau$ into $\mu$ is 18\%, we obtain a 13\% probability of a $\nu_\tau$ producing a muon (and thus $\sim 87$\% of $\nu_\tau$ interactions ending into cascade-like events).}. 
These have to be intended as the topology fluxes impacting the detectors, thus corresponding to the signal event rates before any trigger or selection is applied to the dataset, whose relative efficiencies will impact the final abundance of each event sample in the different detectors.
The resulting maps are then smeared according to instrument angular resolutions: for KM3NeT, we considered upgoing tracks and cascades, as simulated in \cite{km3Astro} for the final ARCA detector configuration, providing the median uncertainty of the reconstructed neutrino arrival direction as a function of the neutrino energy $R_{50}(E_\nu)$. For IceCube cascades, we consider the angular resolution of DNN-reconstructed events, directly resulting from the Galactic Plane analysis \cite{IceCubeScience}. In turn, for the IceCube track-like event selection from the Galactic Plane, we adopt ESTES angular resolution as in the latest Southern sky search \cite{ICestes}. 

The energy-dependent smoothing of the predicted maps is performed by assuming a 2D symmetric Gaussian Point Spread Function (PSF), with an energy-dependent standard deviation $\sigma_{\rm{PSF}}(E_\nu)$, which is derived from the median angular uncertainty via the following relation:
\begin{equation}
    \sigma_{\rm{PSF}}(E_\nu)=\frac{R_{50}(E_\nu)}{\sqrt{2 \ln(2)}}.
\end{equation}

\begin{figure}
    \centering 
    \includegraphics[width=\textwidth]{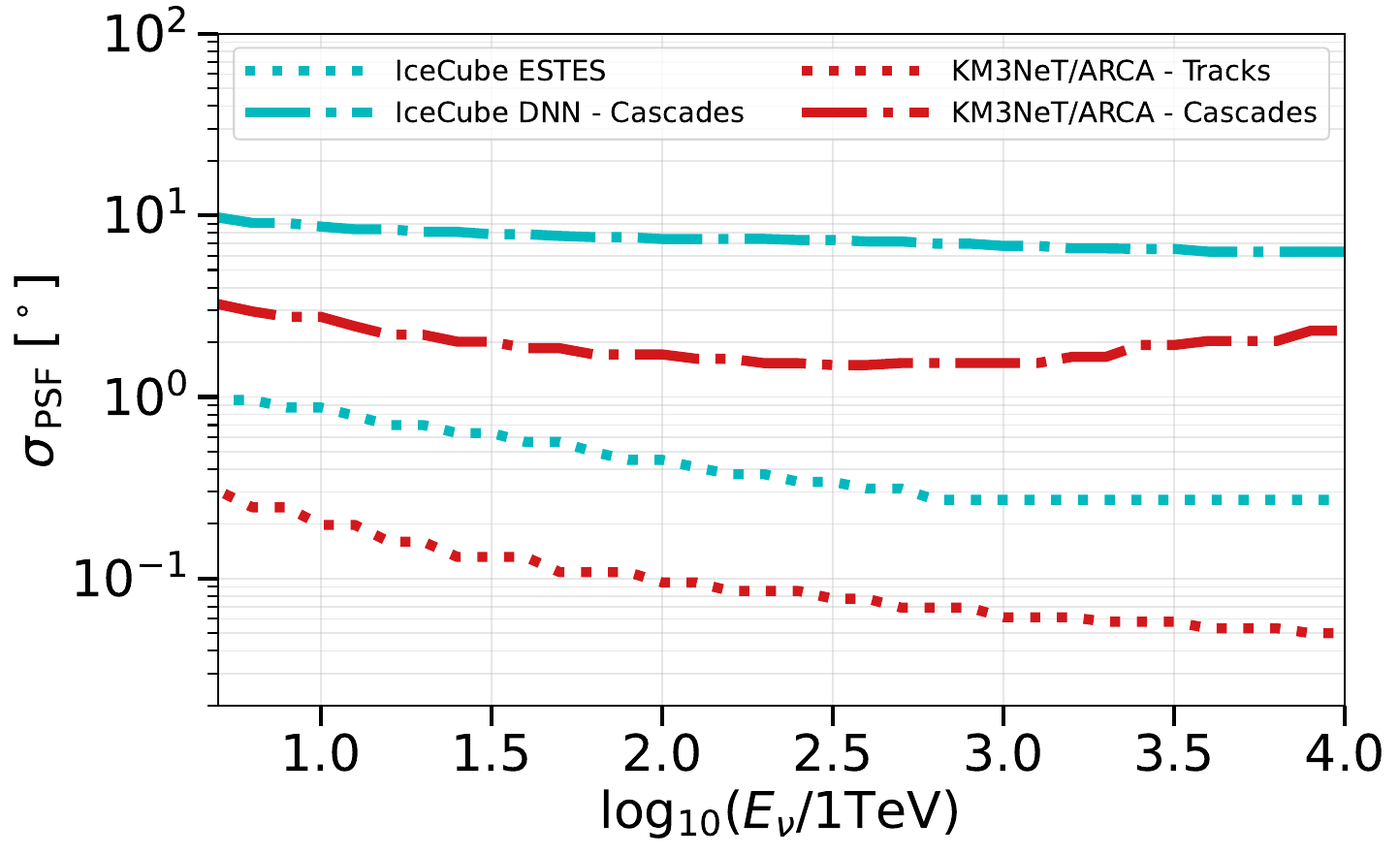}
    \caption{PSFs with track-like (dot-dashed lines) and cascade-like (dotted lines) events of IceCube (light blue lines) and KM3NeT/ARCA (red lines) for Southern Sky observations. IceCube ESTES \cite{ICestes} are adopted for the track sample, while DNN-cascades directly refer to the Galactic Plane analysis \cite{IceCubeScience}. KM3NeT/ARCA tracks and cascades refer to the upgoing selection in the final two building block configuration \cite{km3Astro}.}
    \label{fig:PSFs}
\end{figure}

The latter is reported in Fig.~\ref{fig:PSFs} for each instrument and event sample, while the expected template maps in each channel are shown in  Fig.~\ref{fig:nuTemplates2}, integrated above neutrino energies of 10~TeV, after applying the energy-dependent PSF convolution. For the sake of simplicity, we only show the Kraichnan diffusion scenario as a representative case for the YMSC component, as well as the CR-induced neutrinos for the IceTop \emph{knee} measurement with no spatial-dependent spectral slope.
The top two panels display the expectations for IceCube, with ESTES tracks (first panel from the top) and DNN-cascades (second panel from the top), while the two bottom panels show the corresponding maps for KM3NeT, for upgoing tracks (third panel from the top) and cascades (fourth panel from the top).
The key potential of the track-like event channel in resolving the morphology of the Galactic neutrino emission is evident from these panels and from their comparison with current IceCube cascade measurements.

\section{Comparison of CR-induced neutrino emission with different target gas maps}
\label{app:C}
In order to assess the systematic uncertainties arising from adopting a specific gas distribution in the ISM, we here consider as an alternative Galactic gas model the one presented in \cite{Soding_2025}, providing a three-dimensional reconstruction of the distributions of atomic and molecular gas in the Milky Way. 
By using a Bayesian modelling framework, the observed emission line is modelled via the solution of the radiative transfer equation, including absorption effects and linking the data to the underlying 3D fields of gas density, velocity, and line width. 
The spatial distribution is then recovered through correlated Gaussian processes, which permit coherence across different lines of sight, thus recovering the overall 3D structure of the gas. The conversion from observables to physical gas densities assumes a fixed HI spin temperature of $200\,\mathrm{K}$. For the molecular component, the CO emission is converted to H$_2$ density by adopting a constant conversion factor $X_{\mathrm{CO}} = 2 \times 10^{20}\,\mathrm{cm}^{-2}\,(\mathrm{K\,km\,s^{-1}})^{-1}$.

\begin{figure}
    \centering \includegraphics[width=\textwidth]{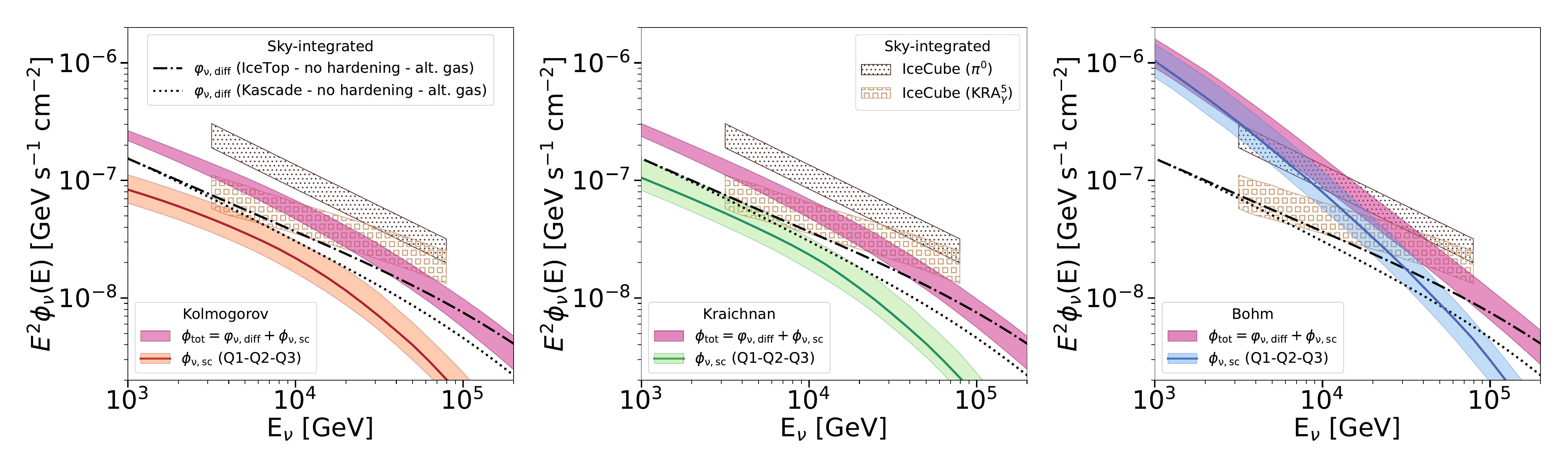}
    \caption{Same as top panels of Fig.~\ref{fig:nuFluxes}, but the CR-induced diffuse emission has been computed with the gas model provided by \cite{Soding_2025}.}
    \label{fig:nuSpectra_NewGas}
\end{figure}

\begin{figure}
    \centering \includegraphics[width=\textwidth]{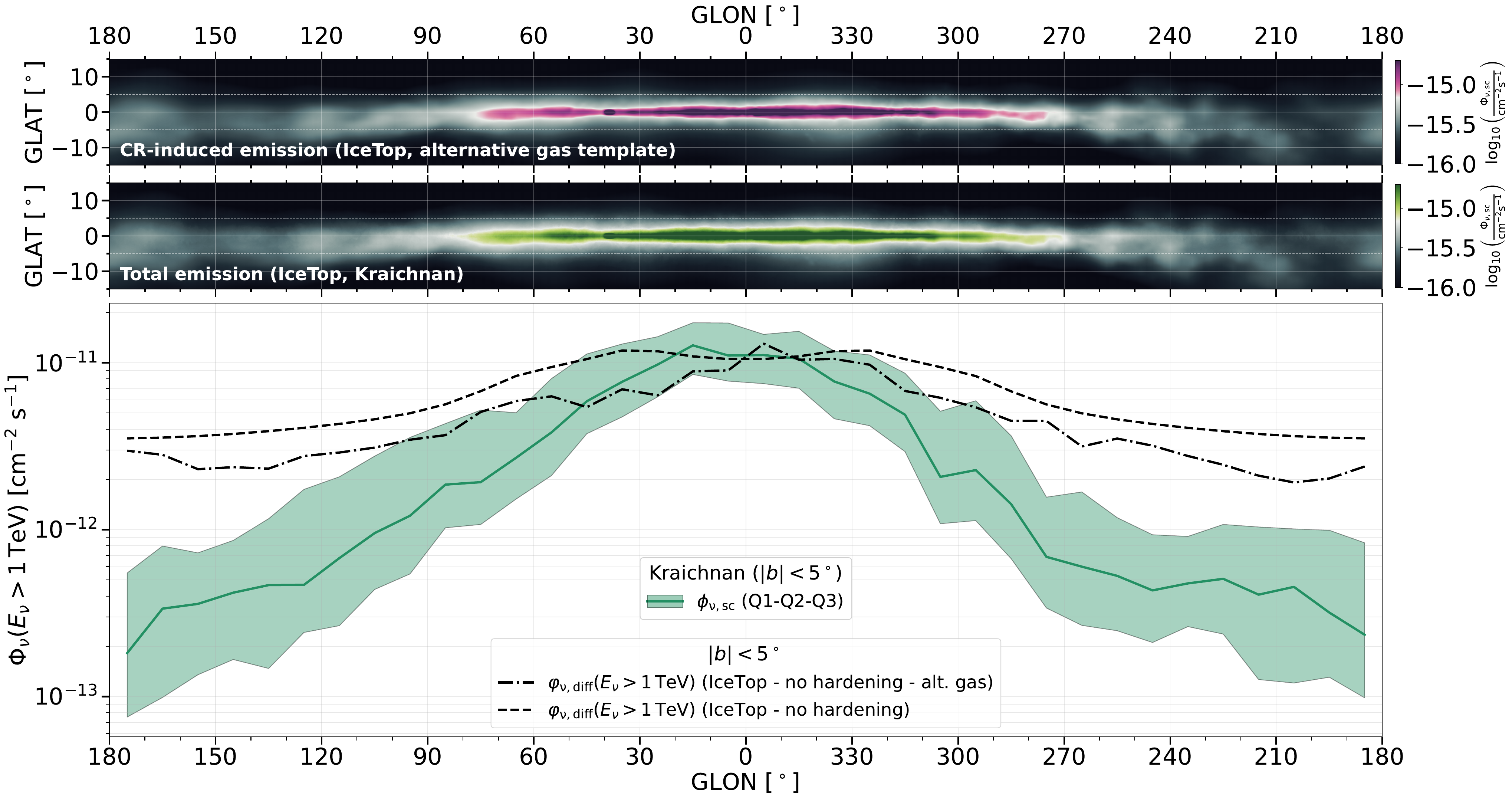}
    \caption{Predicted integral ($E_\nu > 1\,\text{TeV}$) neutrino fluxes, calculated with the alternative gas template from \cite{Soding_2025}. 
    \textit{Top panel:} Emission map of the CR-induced diffuse neutrino flux. \textit{Middle panel:} Total neutrino emission map, i.e. sum of the CR-induced diffuse emission (top panel) and the median emission from YMSCs calculated under the Kraichnan diffusion scenario.
    \textit{Bottom panel:} Longitudinal profile integrated over Galactic latitudes $|b| < 5^\circ$. The dashed black line shows the CR-induced diffuse neutrino emission using the \texttt{GALPROP} gas model, while the dot-dashed black line corresponds to the alternative gas template from \cite{Soding_2025}. The solid green line shows the median integrated flux from stellar clusters ($Q_2$), with the shaded green band representing the interval between the first ($Q_1$) and third ($Q_3$) quartiles of the cluster flux distribution under the Kraichnan diffusion model.}
    \label{fig:nuTemplates_nuProfile_NewGas}
\end{figure}

The alternative gas distribution model results in a different CR-induced neutrino emission from the Galactic Plane, which is shown in Fig.~\ref{fig:nuSpectra_NewGas} as an all sky-integrated neutrino flux, on top of the YMSC neutrino contribution and compared with IceCube measurements.
The normalization of the CR-induced component obtained using the gas model from \cite{Soding_2025} is systematically lower, by about 30\%, than that obtained with the reference gas distribution from \texttt{GALPROP}. 
As a result, the total flux is reduced accordingly, while remaining consistent with the normalization of the KRA$_\gamma$ and $\pi^{0}$ models fitted to the IceCube data (due to the large low-energy discrepancy among them). We further note that the reduction in the CR-induced flux enhances the relative importance of the contribution from YMSCs. \\
Finally,  Fig.~\ref{fig:nuTemplates_nuProfile_NewGas} shows the energy integrated neutrino emission along the Galactic Plane, both for the purely diffuse CR-induced contribution and for the total emission (including YMSCs in the Kraichnan scenario). The bottom panel shows the longitudinal profile integrated over $|b| < 5^\circ$, where it is evident that the contribution from star clusters becomes particularly significant towards the Galactic Center, confirming the results previously obtained with the reference gas model.

\section{Comparison with Galactic diffuse $\gamma$-ray emission}
\label{sec:appA}
In addition to neutrinos, hadronic interactions also produce $\gamma$-rays via the decay of neutral pions. When individual clusters are not resolved, their emitted $\gamma$-ray radiation effectively merges with the diffuse emission of the Galactic Plane, hereby contributing a potentially non-negligible fraction to the total flux.

\begin{figure}
    \centering
    \includegraphics[width=\textwidth]{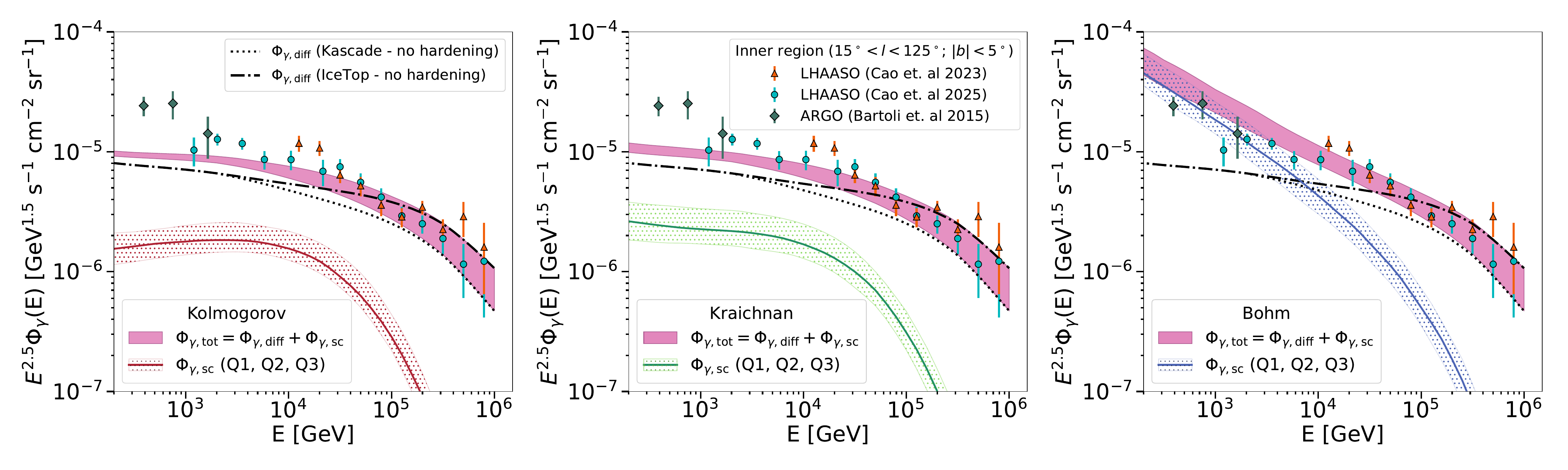}
    \caption{Predicted $\gamma$-ray energy fluxes in the inner Galactic Plane ($15^\circ<l<125^\circ$; $|b|<5^\circ$): the individual contribution by YMSCs and by the CR sea is shown separately, as described in the legend. Diffusion domain is assumed to be Kolmogorov in the left panel, Kraichnan in the middle panel, and Bohm in the right panel. Data points are taken from LHAASO \citep{Cao2023, Cao2025} and ARGO \citep{Bartoli2015} 
    observations.}
    \label{fig:diffuseG}
\end{figure}

To ensure that our predictions for the diffuse neutrino emission are consistent with current observations of the Galactic diffuse $\gamma$-ray emission (GDE), we estimate the contribution of unresolved YMSCs to the GDE and compare it with recent measurements by the ARGO \citep{Bartoli2015} and LHAASO \citep{Cao2023, Cao2025} experiments. Since LHAASO measurements of the GDE are provided after masking resolved sources and objects in existing $\gamma$-ray catalogs, we use a similar procedure to our model predictions. More precisely, we follow the same approach of \cite{Menchiari25}, which consists in applying to the estimated diffuse emission a mask defined as the union of two components:
\begin{itemize}
    \item The first component is a mask that excludes the Local Arm (a disk centred on [$l = 73.5^\circ$, $b=0^\circ$]) and the inner Galactic Plane ($l \leq 70^\circ$, $|b|\leq 1.5^\circ$). This emulates the removal of Galactic sources, such as PWNe, which are not included in our Galaxy model and are known to be concentrated in the first and fourth Galactic Quadrants. 
    \item The second component is a mask that removes all sky regions where the predicted emission from YMSCs at 100 TeV has a statistical significance larger than 5$\sigma$ compared to the measured GDE \citep[see Appendix~B in Ref.][for a detailed definition]{Menchiari25}. This ensures that clusters bright enough to be individually detectable by LHAASO are excluded from the diffuse calculation. 
\end{itemize}
On top of the contribution from unresolved YMSCs, we include the CR-induced $\gamma$-ray flux (similarly masked): this is computed by using Eq.~\eqref{eq:PhiNu_sea} and replacing the neutrino production cross section with that of $\gamma$-ray production \citep{Kachelriess:2022khq}. For this calculation, we adopt the reference gas distribution provided by the \texttt{GALPROP} code \citep{Galprop}. \\
The resulting spectrum is shown in Fig.~\ref{fig:diffuseG}: the comparison with ARGO and LHAASO data shows that the predicted $\gamma$-ray flux in the Kolmogorov and Kraichnan diffusion scenarios remains consistent with the observed GDE, not exceeding the measured emission across the explored energy range. In contrast, the Bohm diffusion case results in a significantly harder spectrum, which slightly overshoots the data at a few TeV. 
This behaviour is in line with previous findings, where the Bohm scenario was already disfavoured due to the excessively large number of YMSCs predicted to be individually detectable by LHAASO \cite{Menchiari25}. 
We note, however, that the flux obtained in this work differs from the one presented in \cite{Menchiari25} for two main reasons: firstly, we improved the modelling of the target gas distribution, considering a more accurate gas profile; secondly, we included the contribution of SNRs. The introduction of a non-constant target density has the effect of lowering the flux intensity of the $\gamma$-ray emission with respect to previous predictions.
Moreover, the CR-induced emission used in this work is calculated under different assumptions with respect to those in \cite{Menchiari25}. 
Additionally, both components are calculated using a more recent parametrization for the $pp$ cross-section (AAFRAG \citep{Kachelriess:2022khq}) rather than SIBYLL \citep{Kafexhiu:2014cua}.

Overall, these results indicate that the population of unresolved YMSCs considered in this work provides a contribution to the GDE that is at most of the order of a few tens of percent in the $\sim$TeV energy range and smaller at lower and higher energies.  
The discrepancy at a few TeV with the data could become slightly larger if we consider the alternative gas model \citep{Soding_2025} in the calculation of the diffuse emission from the CR sea: in fact, it produces approximately $\sim 25 \%$ ($\sim 40 \%$) less signal in the inner (outer region).
This leaves room for additional unresolved source populations.
It has already been shown that the excess cannot be explained as a propagation effect, for example through the hardening of the CR spectral index toward the Galactic centre \citep{Vecchiotti:2024kkz, Luque_2025}.
Recent studies of the unresolved-source contribution to the LHAASO diffuse emission suggest that additional components may still be required, particularly in the inner Galactic region \citep{He_2025}.
Although the contribution of PWNe to the LHAASO diffuse emission appears to be subdominant \citep{Kaci:2024lwx,Vecchiotti:2024kkz}, extended sources, such as TeV halos \citep{Dekker:2023six, Yan:2023hpt} and extended cocoon around sources \citep{PhysRevD.111.083040, Ambrosone:2025wxc}, could still provide a significant contribution to the diffuse $\gamma$-ray emission.
Nevertheless, we highlight that it is still possible though that shell fragmentation might increase the density in the bubble, causing an increment of the flux normalization and leading to a higher contribution of YMSCs to the GDE. Similarly, at sub-TeV energies the inclusion of leptonic emission might increase the relative importance of YMSCs. Accurately estimating the leptonic component is, however, not straightforward, since energy losses prevent an analytical solution of the transport equation. Nonetheless, the recent development of publicly available and optimized numerical tools such as \texttt{SAETASS} \citep{SAETASS} will make it possible to systematically evaluate the contribution of this component in future studies as well.

\acknowledgments

The authors thank M.~Lamoreaux for fruitful discussions about the manuscript content. SC gratefully acknowledges support from the “Award Horizon Europe 2025” funding scheme by Sapienza Università di Roma under grant ID AH1251992EC2A31C. SM and RLC acknowledge financial support from the Severo Ochoa grant CEX2021-001131-S funded by MCIN/AEI/ 10.13039/501100011033. 
GM is partially supported by the INAF Theory Grant 2024 {\it Star Clusters As Cosmic Ray Factories II}.
The work of VV is supported by the European Union’s Horizon Europe research and innovation programme under the Marie Skłodowska-Curie grant agreement No. 101208655 (CORNO GRANDE–COnstRaiNing the Origin of Galactic cosmic RAys using $\gamma$-ray and Neutrino Diffuse Emissions).

\bibliographystyle{JHEP}
\bibliography{biblio.bib}

\end{document}